\documentclass[aps,prx,groupedaddress, superscriptaddress,longbibliography,reprint,twocolumn]{revtex4-2}

\usepackage{amsmath}  
\usepackage{amssymb}
\usepackage{amsthm}
\usepackage{bbm}
\usepackage{graphicx}
\usepackage{color}
\usepackage{upgreek}
\usepackage[linkcolor = blue, citecolor = blue, urlcolor = blue, colorlinks = true]{hyperref}

\usepackage{amssymb}
\usepackage{bm}
\usepackage[capitalise]{cleveref}
\usepackage{textcomp}
\usepackage{hyperref}
\usepackage[usenames,dvipsnames]{xcolor}
\usepackage[normalem]{ulem}
\usepackage{wasysym}

\usepackage{floatrow}
\usepackage{wrapfig}
\usepackage{pgfgantt}

\usepackage{mathptmx}
\usepackage{bm}

\usepackage{fancyhdr}
\usepackage{blindtext}
\fancypagestyle{mypagestyle}{
\fancyhf{}
\fancyfoot[C]{\thepage}  

}
\makeatletter
\let\ps@plain\ps@mypagestyle
\makeatother
\begin{document}

\title{Diffusiophoretic transport of colloids and emulsions in complex environments}

\author{Amir A. Pahlavan}%
 \altaffiliation{Corresponding author}
 \email{amir.pahlavan@yale.edu}

\affiliation{%
Mechanical Engineering and Materials Science, Yale University, New Haven, Connecticut 06511, USA\\
}%

\date{\today}

\begin{abstract}
\vspace{-0.5\baselineskip}
Chemical gradients are ubiquitous in porous and crowded environments, including soils, filters, fabrics, tissues, hydrogels, biofilms and living cells. They arise from displacement fronts, dissolution and precipitation, ion exchange, metabolism, root exudation, evaporation, gas dissolution, freeze--thaw cycles and externally imposed chemical treatments. These gradients can drive colloids, macromolecules and emulsion droplets by diffusiophoresis, while simultaneously driving diffusioosmotic flows along confining surfaces. Classical models of colloid transport in porous media emphasize hydrodynamic dispersion, surface interactions, straining, deposition, detachment and filtration. This chapter places diffusiophoresis within that broader transport framework and reviews how porous media generate, stretch, disperse and sustain the solute gradients that drive phoretic motion. We first discuss sources of chemical gradients and the distinction between spreading and mixing, then summarize classical colloid transport, the minimal physicochemical model for diffusiophoresis and diffusioosmosis, and the experimental platforms used to study these effects. Particular emphasis is placed on recent results showing that diffuse solute fronts can enhance phoretic removal from dead-end pores by prolonging the duration of forcing, and that cross-streamline migration within flowing pathways can change macroscopic breakthrough and dispersion by orders of magnitude. We close by discussing emulsion droplets, multiphase flows, confined and living media, and open problems, including the transition from algebraic mixing in two-dimensional micromodels to chaotic mixing in three-dimensional porous media.
\end{abstract}

\maketitle

This chapter is part of the book {\it Diffusiophoresis and Diffusioosmosis: Theory, Experiment and Applications}, Editors: Guido Bolognesi,  Ankur Gupta, Soft Matter Series of the Royal Society of Chemistry (DOI:10.1039/9781837678136).

\tableofcontents


\section{Motivation and scope}\label{sec:intro}

Porous and crowded materials are chemical landscapes as much as they are hydraulic networks. A fluid moving through soil, rock, a filter, a fabric, a hydrogel, a biofilm, a tissue or a living cell carries suspended colloids, macromolecules and emulsion droplets together with salt, pH, surfactant, nutrients, metabolites, dissolved gases and reaction products. These solute fields are created by displacement fronts, dissolution and precipitation, ion exchange, metabolism, root exudation, drying, gas dissolution, freeze--thaw cycles and engineered chemical treatments \citep{Berkowitz16,Li17,Rolle19,Ladd21,Dentz23,Browne24,Sanquer24,LeBorgne26}. Once present, they are stretched, split, delayed and dispersed by the same pore-scale heterogeneity that controls particle residence times. The central premise of this chapter is that these chemical fields should not be treated only as passive tracers or background conditions: their gradients can themselves act as transport fields.

This viewpoint broadens the classical problem of colloid and droplet transport in porous media. Predicting particle motion, retention and release is essential for contaminant spreading and remediation in the subsurface, colloid-facilitated transport, filtration, membrane separation, cleaning, drug delivery through biological matrices and enhanced oil recovery \citep{McCarthy89,Kretzschmar99,Bradford02,Tratnyek06,Kanti06,Klaine08,Bradford08,Weber09,Bradford13,Pokrovsky16,Morales2017,Machado18,Aramideh19,Ouyang20,MacLeod21,Mangal21,Mangal21b,Patino23,Wu23,Rillig24,Pahlavan24,Huang24,Donnell24,Storm24,Skierszkan25}. The established framework emphasizes hydrodynamic dispersion, Brownian motion, interception, straining, deposition, detachment, aggregation and DLVO-type particle--surface interactions \citep{Liu95,Johnson96,Roy97,Torkzaban07,Molnar15,Molnar19,Bizmark20,Mangal22,Darko24}. These mechanisms remain indispensable, but they do not exhaust the role of chemistry: a local concentration modifies surface forces and aggregation, while a concentration gradient can drive directed particle motion and wall slip.

Diffusiophoresis and diffusioosmosis provide the interfacial mechanism for this coupling \citep{Anderson1989,Velegol16,Marbach19,Shim22,Ault24,Shi25}. For dilute electrolytes, the particle velocity is proportional to the gradient of the logarithm of solute concentration; for neutral solutes, concentrated electrolytes and multicomponent solutions, the driving is more generally expressed through chemical-potential gradients or species fluxes. This logarithmic response is important in porous media because the particle responds to relative concentration changes, so weak but persistent gradients can still produce measurable motion. Since solutes diffuse much faster than micron-scale colloids, $D_s\gg D_p$, a solute front can explore a stagnant or weakly connected region on the time scale $L^2/D_s$ while a colloid would require $L^2/D_p$ to do the same. Diffusiophoresis therefore offers a route for converting fast solute diffusion into slow particle displacement.

Much of what is known about these effects comes from microfluidic experiments in which sharp solute gradients are imposed and particle motion is studied in quiescent conditions or in simple flow geometries \citep{Abecassis08,Palacci10,Palacci12,Yadav13,Florea14,Guha15,Chiang14,Shi16,Shin16,Ault17,Nery17,Ault18,Battat19,Gupta19,Wilson20,Gupta20,Singh20,Tan21,Alessio21,Jotkar21,Alessio22,Shim22b,Chakra23,Akdeniz23,Lee23,Migacz23,Liu25,Chakra25}. A porous medium is less controlled. The same flow that transports the particles also stretches, splits and disperses the solute, so the gradients that drive phoresis are heterogeneous, transient and often weak \citep{Villermaux19,Dentz23,LeBorgne26}. Mean speeds in preferential pathways can be $\mathcal{O}(10^2$--$10^3)~\mu\mathrm{m}/\mathrm{s}$, whereas calibrated phoretic velocities are often only $\mathcal{O}(1$--$10)~\mu\mathrm{m}/\mathrm{s}$ \citep{Shin16,Ault24,Alipour26}. Yet porous media also contain broad velocity distributions, preferential pathways, nearly stagnant pockets and long residence-time tails, producing non-Fickian behavior commonly described using continuous-time random walks and multi-rate mass-transfer models \citep{Saffman59,Koch85,Koch88,Sahimi93,Bijeljic04,Berkowitz06,Gouze08a,Gouze08,Borgne08,Neuman09,Borgne10,Dentz11,Anna13,Blunt13,Bordoloi22}. The relevant question is not whether phoresis overwhelms the mean flow, but whether it changes which regions of the velocity field particles sample, how long they remain trapped and where they are deposited or released.

How a flowing porous medium stretches, disperses and mixes a scalar field is therefore the origin of the forcing that drives diffusiophoresis \citep{Jotkar24,Jotkar24b,Pujari26,Li26,Alipour26}. Migration of colloids across streamlines under solute gradients is itself not new; it has been demonstrated in idealized stagnation-point, cellular, chaotic and turbulent flows, where diffusiophoresis can compress or disperse particles relative to the solute and alter suspension mixing \citep{Deseigne14,Mauger16,Raynal18,Raynal19,Volk22,Shukla17}. What has remained open is whether, and by what mechanism, weak transverse migration operates within the flowing backbone of a porous medium, and how it competes with exchange into and out of stagnant regions. This chapter is organized around that shift in viewpoint: diffusiophoresis in porous media is both an exchange mechanism with dead-end pores and a redistribution mechanism within flowing pathways.

Two questions form the quantitative spine of the chapter:
\begin{enumerate}
    \item Does dispersion of solute fronts in porous media weaken diffusiophoretic migration of colloids across dead-end pores, or can diffuse fronts enhance removal by prolonging the duration of forcing?
    \item Is the influence of diffusiophoresis limited to dead-end pores and stagnant pockets, or can weak transverse migration within preferential pathways reshape breakthrough, residence times and macroscopic dispersion?
\end{enumerate}
These questions also explain why the same dimensionless quantities recur throughout the chapter: the phoretic strength relative to solute diffusion, $\Gamma_p/D_s$; the imposed concentration ratio, $c_1/c_0$; the solute P\'eclet number; the mobility ratio between particles and walls; and the statistics of the pore-scale velocity field.

We first survey the sources of chemical gradients in porous and crowded environments, distinguish solute spreading from mixing, summarize classical colloid transport, introduce the minimal physicochemical model for diffusiophoresis and diffusioosmosis, and describe the microfluidic platforms used to isolate these mechanisms. The central sections then review two recent porous-media results: diffuse fronts can enhance phoretic removal from dead-end pores, and cross-streamline migration within flowing pathways can change macroscopic dispersion by orders of magnitude \citep{Jotkar24,Jotkar24b,Li26,Pujari26,Alipour26}. The later sections extend the same framework to emulsion droplets, whose deformable and surfactant-laden interfaces introduce Marangoni stresses, aggregation, coalescence, mobility reversal and pore blockage, and to confined or living media such as fabrics, hydrogels, biofilms, bacteria, microtubules and biomolecular condensates \citep{Yang18,Park21,McKenzie22,Duong26,Shin18,Doan21,Sambamoorthy23,Somasundar23,Ramm21,Alessio23,Doan24,Hafner24,Shim24}. We finish the chapter by highlighting the outlook and open problems that remain when one moves from idealized microfluidic geometries to three-dimensional, reactive, multiphase and living materials.

\section{Sources of chemical gradients in porous environments}\label{sec:gradient_sources}

Here, \emph{porous environments} refers to several classes of materials that appear repeatedly in this chapter: granular soils and sediments; sandstone, carbonate and fractured or Karst aquifers; engineered filters, membranes and fabrics; soft hydrated matrices such as hydrogels, mucus, extracellular matrix, and tissues and biofilms \citep{Bear88,Blunt02,Molnar15,Molnar19,Shin18,Doan21,Somasundar23,Dentz23,Yang2026Multiphase}. These materials differ greatly in stiffness, pore size and surface chemistry, but they share two features that are central for diffusiophoresis: large internal surface area and strong spatial heterogeneity in flow, composition and residence time. The relevant scalar fields can include salt, pH, surfactant, nutrients, metabolites, redox-active species, dissolved gases, reaction products and macromolecules, arising from hydrologic mixing, reactive transport, biological activity, drying, gas dissolution and engineered chemical treatments \citep{Berkowitz16,Rolle19,Valocchi19,Marbach19,Dentz23,LeBorgne26,Browne24,Sanquer24}. For dilute electrolytes the simplest phoretic law is logarithmic, $\mathbf{u}_{\mathrm{dp}}=\Gamma_p\nabla\ln c$; for neutral solutes, concentrated electrolytes or multicomponent solutions, the driving is more generally expressed through chemical-potential gradients or species fluxes, and the mobility can have either sign (Section~\ref{sec:fundamentals}). Thus the key issue is not simply whether a large absolute concentration difference exists, but whether pore-scale processes create a relative gradient that persists long enough to move a particle. As a scale, a tenfold concentration change spread over $10$--$100~\mu$m with $\Gamma_p=10^{-10}$--$10^{-9}~\mathrm{m^2\,s^{-1}}$ gives phoretic speeds of order $1$--$10~\mu$m/s, a range measurable in microfluidic devices and comparable to Brownian escape or to local advective speeds in slow regions.

 \begin{figure*}[htb!]
    \centering
    \includegraphics[width=1\textwidth]{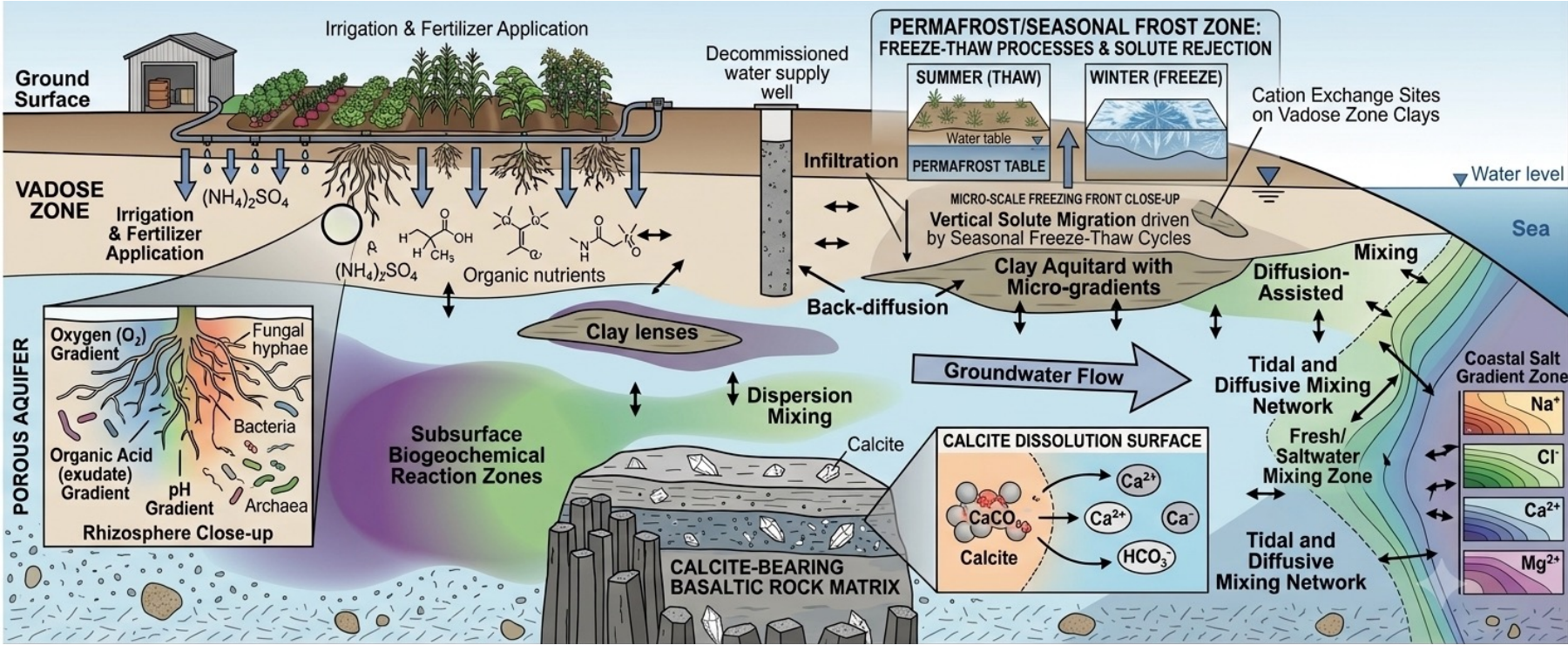}
    \caption{Different sources of chemical gradients in subsurface environments, from irrigation and fertilizer application in agriculture to biogeochemical reactions, freeze-thaw cycles, and salinity gradients in coastal zones.}
    \label{Fig1}
\end{figure*}

\noindent \textbf{Displacement fronts:} are representative of many relevant scenarios, including surfactant rinsing of fabrics, low- or high-salinity water flooding, groundwater recharge, irrigation, fertilizer or pesticide pulses, and engineered remediation, where a resident solution of concentration $c_0$ is displaced by an invading solution of concentration $c_1$ \citep{Shin18,Park21,Muller07,Li17,Tratnyek06,Alipour26,Li26}. Such fronts are appealing precisely because the initial and final states are known, even though the pore-scale gradient field that develops between them is not directly controlled. In a porous medium, the solute front is stretched, split, delayed in slow zones and dispersed by the velocity field, producing a heterogeneous field of $\nabla\ln c$ (Fig.~\ref{Fig1}). This class of problems provides the experimental foundation for the dead-end-pore and flowing-backbone studies discussed later in the chapter.

\noindent \textbf{Self-sustained gradients:} mineral dissolution produces local gradients of ions, alkalinity and pH \citep{Molins21,Kar16}; precipitation and carbon mineralization consume species and can reshape both local chemistry and pore geometry \citep{Kelemen12,Matter16,Yang24b}; redox reactions, acid--base fronts and ion exchange maintain composition gradients over distances from individual pores to aquifers \citep{Berkowitz16,Rolle19,Valocchi19,Dentz23,Beinlich20}. Such gradients are usually discussed as controls on reaction rates and alteration of the solid matrix, but they can also act directly on suspended colloids. A recent calcite-dissolution micromodel illustrates the feedback: gradients generated by dissolution drove colloids toward the reacting mineral, where aggregation formed a passivating layer and slowed the dissolution process itself \citep{Roman25}.

\textbf{In many energy and environmental technologies} analogous gradients emerge. Enhanced oil recovery mixes seawater, low-salinity brines, surfactants, polymers and alkaline solutions with resident reservoir fluids, producing salinity, surfactant and pH gradients in a multiphase pore space \citep{thomas2008,Blunt02,Park21}. Geological carbon storage and carbon mineralization generate gradients of dissolved CO$_2$, carbonate species, pH, divalent cations and saturation state as injected fluids mix and react with rock \citep{Huppert2014,Fu15,Matter16,Kelemen12}. Underground hydrogen storage, geothermal circulation and in situ remediation similarly involve injected fluids that mix with native brines, react with minerals or microbes, and alter local redox state, pH and ionic strength \citep{Heinemann21,Tratnyek06,Zhang19}. During evaporation or drying of partially saturated media, and during dissolution of injected CO$_2$ or H$_2$ into resident brine, solutes are concentrated, can exceed saturation and can precipitate, generating salinity and ionic gradients local to a receding interface or reacting front \citep{Matter16,rees-zimmerman2021,Heinemann21}. In these settings concentration fields affect transport in two distinct ways: their gradients drive phoretic motion, while their local values modify wettability, interfacial tension, aggregation and permeability through salinity-dependent surface forces, surfactant adsorption/desorption, mineral precipitation/dissolution, and particle deposition or clogging \citep{Blunt02,Lee16,Park21,Yang24b,Bizmark20,Roman25}.

\textbf{A third class consists of \emph{maintained gradients}}, where boundary conditions replenish two solutions and the scalar field approaches a quasi-steady state rather than a single transient displacement. Examples include parallel streams in microfluidic devices, hydrothermal and alkaline-vent systems in which reduced alkaline fluids mix with seawater, estuaries and coastal aquifers in which tidal forcing maintains salinity gradients, and river--groundwater exchange zones where freshwater and saline or geochemically distinct waters meet \citep{Abecassis08,Martin08,Ianeselli23,Zhang17,Wang22,Berkowitz16}. These settings are important because diffusiophoretic migration can persist as long as the gradient is maintained, and the relevant comparison is then between the phoretic drift time and the advective residence time in the gradient zone.

\textbf{Biological porous materials} provide a different set of sources and constraints. In extracellular matrix, tumors, mucus, collagen gels and biofilms, gradients of ions, nutrients, oxygen, metabolites and waste products arise from cellular metabolism, diffusion limitation and external flows through matrices that are crowded, tortuous and chemically heterogeneous \citep{Heldin04,Wiig12,Nicholson2017,Carrel18,Kurz23}. Biofilms are especially clear examples because they are simultaneously porous media and chemical reactors: the extracellular polymeric matrix restricts transport, while embedded cells consume nutrients, produce metabolites and remodel the pore space \citep{Carrel18,Kurz23,Scheidweiler19,Somasundar23}. In this context diffusiophoresis can be viewed as a potential delivery or removal mechanism: imposed electrolyte gradients have been shown to drive particles into bacterial biofilms, and collagen-gel experiments demonstrate that phoretic transport can survive strong confinement \citep{Doan21,Somasundar23,Chen26}.

\textbf{The crowded interior of living cells and reconstituted cellular systems} is another setting, where the focus is on membrane-bound cargo, microtubules, biomolecular condensates and phase-separated droplets moving in gradients of proteins, ions, fuel, waste or other macromolecules \citep{Sear19,Ramm21,Burkart22,Doan24,Hafner24,Alessio23,Jambon24,Shim24}. Experiments and theory on these systems show that chemical gradients can couple to pattern formation, phase separation and intracellular organization \citep{Ramm21,Doan24,Hafner24,Alessio23,Shim24}. These systems show how the same nonequilibrium transport principle appears in soft, crowded and chemically active environments.

\section[Mixing in porous-media flows]{Mixing, spreading and scalar-gradient generation in porous-media flows}\label{sec:mixing}

\begin{figure*}[htb!]
    \centering
    \includegraphics[width=1\textwidth]{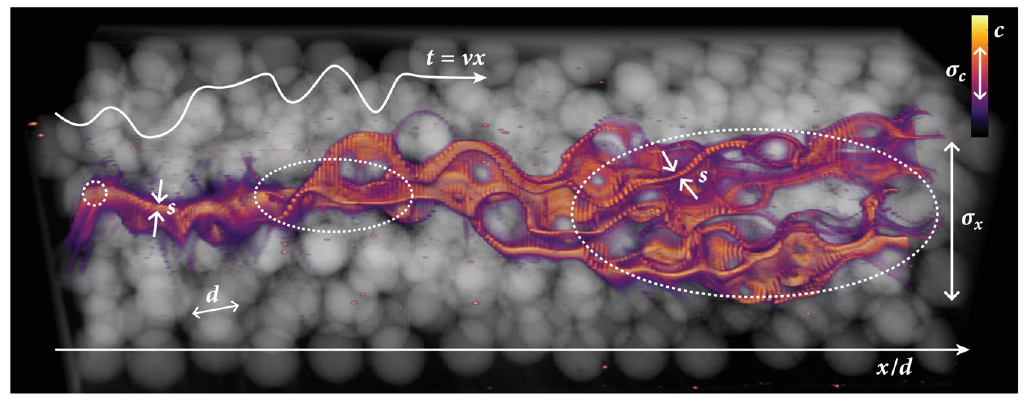}
    \caption{Spreading and mixing of a solute plume in a three-dimensional porous bead pack: stretching, mixing and concentration gradients evolve via distinct mechanisms. Adapted from ref.~\citep{LeBorgne26} with permission.}
    \label{Fig2}
\end{figure*}

The fate of a colloidal suspension is inseparable from the fate of the chemical field that moves it. The same pore-scale flow that carries colloids simultaneously deforms, disperses and mixes the solute, continuously reshaping the gradients on which phoresis feeds. The scalar problem therefore provides the bridge between porous-media transport and diffusiophoresis. Three ideas are especially useful here: hydrodynamic dispersion describes spreading of the mean plume, mixing describes the weakening of concentration contrasts, and fluid deformation creates the pore-scale gradients before diffusion ultimately smooths them (Fig.~\ref{Fig2}).

Dispersion in porous media has a long history that spans fluid mechanics and hydrology, from Taylor's analysis of shear dispersion in a tube and early pore-scale theories by Saffman and others, to continuum and stochastic descriptions of dispersion and macrodispersion in fixed beds, laboratory columns and aquifers \citep{Taylor53,Josselin58,Saffman59,Bear72,Pfannkuch63,Scheidegger61,Gelhar83,Dagan89,Brenner80,Koch85,Koch88,Koch89,Bijeljic04,Dentz18,Dentz23}. In that tradition, dispersion refers to the growth of the plume footprint or to the effective tensor that relates the average solute flux to the gradient of an averaged concentration. Mixing, by contrast, refers to homogenization: the decay of concentration variance and the erosion of concentration contrasts \citep{Villermaux19,Dentz23,LeBorgne26}. The distinction matters for reactions and phoretic transport. Two solute plumes with similar macroscopic dispersion can contain very different pore-scale gradient fields, and therefore can drive very different colloid trajectories.

At the pore scale, a conservative solute concentration $c(\mathbf{x},t)$ in an incompressible liquid satisfies
\begin{equation}
    \frac{\partial c}{\partial t}+\mathbf{u}\cdot\nabla c=D_s \nabla^2 c,\qquad \nabla\cdot\mathbf{u}=0,
    \label{eq:advection_diffusion_solute}
\end{equation}
with no-flux boundary conditions at solid surfaces. At larger scales, this equation is commonly replaced by an advection--dispersion equation in which pore-scale velocity fluctuations are represented by an effective dispersion tensor \citep{Saffman59,Brenner80,Koch85,Dentz18,Dentz23}. That representation is powerful for predicting plume spreading and breakthrough, but it does not preserve the local gradients that control reaction rates and phoretic velocities. The limitation is familiar from reactive transport, where reaction rates depend on the overlap and covariance of reactants rather than only on mean concentrations \citep{Gramling02,Anna14b,Berkowitz16,Rolle19,Valocchi19}. Diffusiophoresis has the same closure problem because averaging $\nabla\ln c$ is not equivalent to taking the logarithmic gradient of an averaged concentration.

The physical mechanism for gradient generation is fluid deformation. A solute interface carried through a pore network is stretched by velocity gradients and compressed transversely. Molecular diffusion then acts across the narrowed lamellae, so stretching makes diffusion effective by reducing the length over which it must smooth concentration differences. In a lamellar description, the strip thickness $s(t)$ is set by the integrated stretching history along trajectories, and the competition between advective compression and diffusion sets a Batchelor-type scale below which gradients cannot sharpen further (Fig.~\ref{Fig2}) \citep{Ranz1979,VillermauxDuplat2003,Borgne13,Borgne15,Villermaux19,Dentz23}. As neighbouring lamellae thin and overlap, their aggregation governs the late-time concentration probability distribution and the decay of variance. For diffusiophoresis, the same mechanisms determine the magnitude, orientation and persistence of $\nabla\ln c$. 

Porous media flows contain fast preferential pathways, slow regions near solid surfaces and nearly stagnant dead-end pores. Solute carried through fast pathways arrives early, while solute trapped in slow regions is delayed and released gradually. The resulting breakthrough curves often deviate from Gaussian/Fickian behavior before the asymptotic hydrodynamic-dispersion regime is reached \citep{Saffman59,Berkowitz06,Neuman09,Dentz18,Dentz23}. For passive solutes this behavior is described using residence-time distributions, continuous-time random walks or mobile--immobile exchange \citep{Berkowitz06,Dentz11,Bordoloi22}. For diffusiophoretic particles, the same heterogeneity controls where and when gradients are present. Therefore, slow regions are not only particle traps, but are reservoirs of solute history that can sustain gradients after the mean front has passed.

Finally, dimensionality controls how strongly the flow can stretch material elements. Quasi-two-dimensional micromodels are invaluable because they isolate dead-end exchange, cross-streamline migration and geometric disorder, but steady two-dimensional incompressible flows stretch material lines algebraically. In three-dimensional porous media, steady Stokes flow can be chaotic: streamlines braid around grains and throats, material surfaces stretch exponentially, and sharp scalar gradients can persist over long distances \citep{Lester13,Lester16,Turuban18,Turuban19,Heyman20,Souzy:2020aa,LeBorgne26}. We keep this distinction in mind here but treat its consequences for diffusiophoresis as an open problem in the outlook (Section~\ref{sec:outlook}).

\section{Classical colloid transport in porous media}\label{sec:classical_colloid}

\begin{figure*}[htb!]
    \centering
    \includegraphics[width=1\textwidth]{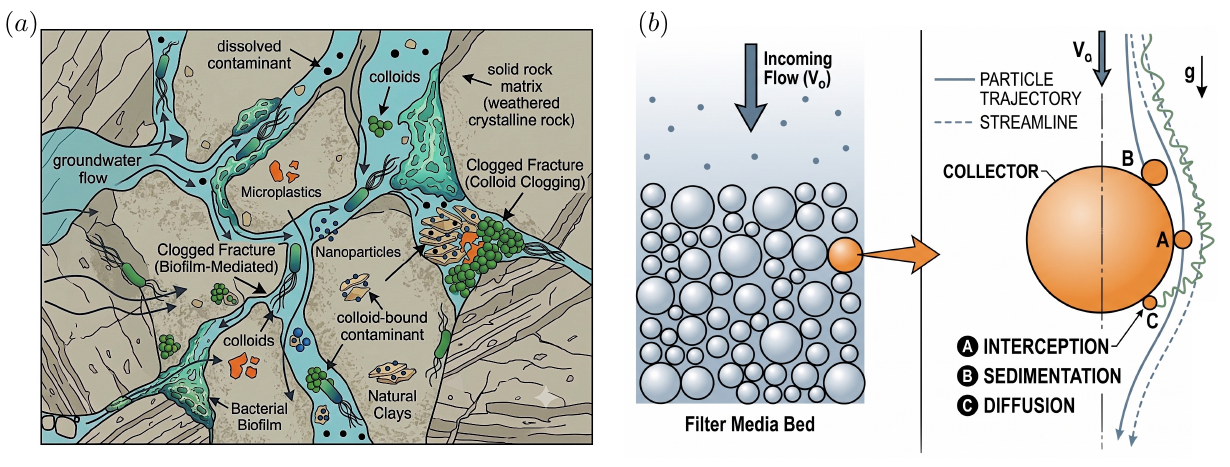}
    \caption{(a) Transport of colloids, contaminants, microplastics, pathogens, and bacteria in subsurface environments. Colloid deposition or bacterial biofilms reshape the transport pathways.  (b) Colloid filtration theory describes deposition as an interplay between DLVO-type interactions, interception, sedimentation and diffusion. Panel (b) adapted from ref.~\citep{Yao1971} with permission}
    \label{Fig3}
\end{figure*}

Before introducing diffusiophoresis, we summarize the classical picture of colloid transport in porous media. This literature spans environmental engineering, hydrogeology, filtration, membrane science, petroleum engineering, medicine and colloid science \citep{Ryan96,Kretzschmar99,Molnar15,Molnar19}. It describes colloid transport through hydrodynamic trajectories, Brownian motion, interception and straining, surface forces, aggregation, attachment and detachment (Fig.~\ref{Fig3}). Diffusiophoresis should therefore be viewed as an additional chemically-driven drift embedded within this framework.

\subsection{Continuum description and the retention closure}\label{sec:colloid_continuum}
At the continuum scale, a suspended colloid concentration $n(\mathbf{x},t)$ is often modeled by an advection--dispersion--retention equation:
\begin{equation}
    \phi\frac{\partial n}{\partial t} + \rho_b\frac{\partial s}{\partial t}
    = -\nabla\cdot(\mathbf{u} n) + \nabla\cdot(\mathbf{D}\nabla n),
    \label{eq:classical_adr}
\end{equation}
where $\phi$ is porosity, $\rho_b$ is bulk density, $s$ is the retained concentration per mass of solid, $\mathbf{u}$ is the Darcy or pore-scale averaged velocity and $\mathbf{D}$ is a dispersion tensor. Closure requires a law for $s$, most simply $\partial s/\partial t=k_a n-k_d s$, and more generally terms for blocking, ripening, straining, release, finite retention capacity or depth-dependent retention \citep{Bradford2003,Johnson1995,Bradford2012release,Molnar15,Molnar19}. Breakthrough curves and retained profiles are then used to infer effective parameters \citep{Molnar15,Molnar19}. For steady spatially uniform deposition, Eq.~\eqref{eq:classical_adr} reduces to $n(x)=n_0\exp(-kx/U)$, the classical log-linear attenuation used to estimate travel distances and interpret column data \citep{Yao1971,Elimelech1995}. Microfluidic and three-dimensional imaging show why this is only a baseline: deposition and erosion are intermittent, local clogging redirects flow, and aggregates can be released in bursts \citep{Bizmark20,Gerber2019,Wu23,Wu24}.

\subsection{Colloid filtration theory and the single-collector efficiency}\label{sec:cft}
Colloid filtration theory (CFT) links the continuum deposition rate to pore-scale capture. Deposition is decomposed into \emph{transport} to a grain, quantified by the single-collector contact efficiency $\eta_0$, and \emph{attachment} after contact, quantified by the attachment efficiency $\alpha$ \citep{Yao1971}. A collector-scale mass balance gives $k=\tfrac{3(1-\phi)}{2d_c}\,v\,\alpha\eta_0$, where $d_c$ is collector diameter and $v$ is pore velocity. Brownian diffusion, interception and gravitational sedimentation contribute to $\eta_0$ and generate the characteristic ``U''-shaped dependence of clean-bed removal on particle size. The classical calculation uses Happel's sphere-in-cell flow field \citep{Happel1958}; later work incorporated hydrodynamic retardation, retarded van der Waals attraction, lattice-Boltzmann simulations and alternative collector geometries \citep{Rajagopalan1976,Tufenkji04,Long2009,Ma2009,Nelson2011}. The central assumption is that each collector removes the same constant fraction $\eta=\alpha\eta_0$, so retention decays log-linearly with depth. Under unfavorable or heterogeneous surface conditions, column and micromodel studies often show hyperexponential, nonmonotonic or history-dependent profiles instead \citep{Tufenkji2003,Tufenkji2004dev,Tufenkji2005breakdown,Molnar15,Bizmark20}.

\subsection{Surface interactions, attachment and anomalous retention}\label{sec:dlvo}\label{sec:breakdown}
Attachment is controlled by nanometer-scale colloid--collector interactions. DLVO theory sums attractive van der Waals and electrostatic contributions to the interaction energy \citep{Derjaguin1941,Verwey1948}. Because many natural colloids and minerals are negatively charged at neutral pH, electrostatic repulsion often creates an energy barrier whose height depends on ionic strength through the Debye length. Under \emph{favorable} conditions, e.g., high ionic strength or oppositely charged surfaces, the barrier is absent and $\alpha\approx1$; under \emph{unfavorable} conditions, a primary maximum opposes attachment to the primary minimum and particles may instead reside weakly in a secondary minimum \citep{Elimelech90,Ryan96,Hahn2004,Redman2004,Tufenkji2004dev}. This is the interaction-energy language that diffusiophoresis modifies: local concentration changes barrier heights and attachment probabilities, whereas concentration gradients add directed drift.

The CFT--DLVO picture is most reliable for clean beds under favorable conditions. In natural unfavorable systems, attachment efficiencies are often much larger than mean-field DLVO predicts, and retained profiles often deviate from the single-rate exponential form \citep{Elimelech90,Tufenkji2004dev,Molnar15}. Hyperexponential and nonmonotonic profiles imply that the effective deposition rate varies with distance, history or population substructure \citep{Tufenkji2003,LiScheibe2004,LiJohnson2005}. One compact interpretation is a distribution of deposition modes: secondary-minimum association can act in parallel with slower primary-minimum attachment \citep{Tufenkji2004dev,Tufenkji2005breakdown}. Another is surface heterogeneity. Trace oxides, organic coatings and nanoscale charge patches create locally favorable sites on nominally unfavorable grains; as those sites fill, deposition naturally becomes depth- and history-dependent \citep{Song1994,Johnson96,Elimelech00,Molnar15}.

\subsection{Physical retention, release and finite-capacity effects}\label{sec:straining}\label{sec:release}\label{sec:concentration}
Retention also occurs through pore geometry. \emph{Straining} traps colloids in throats too small to pass and depends on colloid-to-grain or colloid-to-throat size ratio \citep{Bradford02,Bradford2003}. Early criteria placed the threshold near a diameter ratio of $\mathcal{O}(0.1)$, but experiments show straining at smaller ratios and sensitivity to velocity, concentration, ionic strength, particle shape and surface chemistry \citep{Sakthivadivel1966,Herzig1970,Matthess1985,Bradford2006straining,Xu2006,Shen2008}. Related mechanisms include wedging at converging grain contacts, bridging across constrictions and clogging transitions that redirect flow and promote further capture \citep{Johnson2007wedging,Ramachandran1999,Bizmark20,Gerber2019,Wu23}. Size exclusion acts in the opposite direction: particles excluded from small pores are routed through larger, faster paths and can break through earlier than a conservative solute \citep{Bradford2003,Auset04}. Chemistry and geometry are coupled because hydrodynamic trajectories can carry weakly attached secondary-minimum particles toward grain contacts, where physical trapping is strongest \citep{Bradford2007,Torkzaban07}.

Deposited colloids may later be released, and the deposition rate itself evolves. Lower ionic strength or higher pH expands double layers, increases repulsion and can mobilize particles from secondary or shallow primary minima \citep{Ryan96,Grolimund1998,Roy97}; the rate then depends on detachment across any residual barrier and diffusion through the near-surface stagnant film \citep{RuckensteinPrieve1976,Kretzschmar99}. As particles accumulate, repulsive deposited particles exclude later arrivals and produce blocking, whereas favorable particle--particle interactions create new collectors and produce ripening \citep{Johnson1995,Bradford2012release,Bradford2015determining}. Both imply finite retention capacity. Thus retention can depend on input concentration: site saturation can reduce fractional retention, while bridging and clogging can increase it, with the dominant trend set by ionic strength and geometry \citep{BradfordBettahar2006,Foppen2005,Bradford2009coupled}. In partially saturated media, air--water interfaces, contact lines and wetting films add retention sites and make transport sensitive to hydrophobicity, air content and wetting--drying history \citep{DeNovio2004,Flury2017,WanWilson1994a,Crist2004}.

\subsection{Coupling to the flow field and colloid-facilitated transport}\label{sec:coupling}
A frontier theme, enabled by microfluidics and three-dimensional imaging, is that colloids do not merely respond to a fixed flow; they can reshape it. Deposition, aggregation and biofilm growth alter permeability, redirect fast and slow paths, and change subsequent transport \citep{Gerber2019,Bizmark20,Miele19}. The coupling is especially strong in two-phase flow, where colloids can clog liquid bridges, concentrate at pendular rings and triple points, fragment connected wetting clusters and change relative permeability \citep{Wu23,Wu24}. Accordingly, pore-network and stochastic models have moved beyond constant efficiency toward history-dependent interception and retention formulations that reproduce hyperexponential and nonmonotonic profiles from a single colloid population \citep{Morales2017,Molnar15,Molnar19}.

When metals, radionuclides or hydrophobic compounds sorb to mobile colloids, the colloidal phase provides a fast pathway that can outrun predictions based on dissolved-phase retardation alone \citep{McCarthy89,Kretzschmar99,Grolimund1996}. Laboratory studies show accelerated metal and radionuclide transport by clays and organic colloids, and field observations confirm that colloid-facilitated transport operates in the subsurface \citep{Saiers1996,Kersting99,Penrose1990}.

\subsection{From classical transport to diffusiophoresis}\label{sec:colloid_bridge}
Classical colloid models also face the spreading-versus-mixing issue discussed in Section~\ref{sec:mixing}. Breakthrough curves are often non-Fickian because particles spend broadly distributed times in slow zones, dead-end pores or surface-associated states; mobile--immobile mass transfer, continuous-time random walks and multi-rate retention provide common descriptions \citep{Berkowitz06,Neuman09,Molnar19,Bordoloi22}. Recent pore-scale experiments and simulations sharpen this picture by showing that vortices, dead-end pores, geometric disorder and particle motility can alter residence-time statistics \citep{Anna21,Scheidweiler24}. Diffusiophoresis enters in two ways. First, solute concentration changes attachment, release and aggregation by changing the interaction landscape. Second, solute gradients add drift. In the absence of retention, a phoretic colloid in a known flow and solute field obeys
\begin{equation}
    \frac{\partial n}{\partial t}+
    \nabla\cdot\left[\left(\mathbf{u}+\Gamma_p\nabla\ln c\right)n\right]
    =D_p\nabla^2 n,
    \label{eq:colloid_dp_transport}
\end{equation}
where $D_p$ is the Brownian diffusivity and $\Gamma_p$ is the diffusiophoretic mobility. A complete model must couple this drift to attachment, detachment, straining, aggregation and evolving surface chemistry. Even before those couplings are fully resolved, Eq.~\eqref{eq:colloid_dp_transport} identifies why phoresis can matter in flow: the added drift is not generally aligned with the pressure-driven velocity. A small transverse phoretic velocity can move particles between slow and fast streamlines, alter residence-time distributions and change macroscopic dispersion. The rest of this chapter examines the consequences of adding this chemically driven term to the classical transport picture.

\section{Fundamentals of diffusiophoresis and diffusioosmosis}\label{sec:fundamentals}

\begin{figure*}[htb!]
    \centering
    \includegraphics[width=0.75\textwidth]{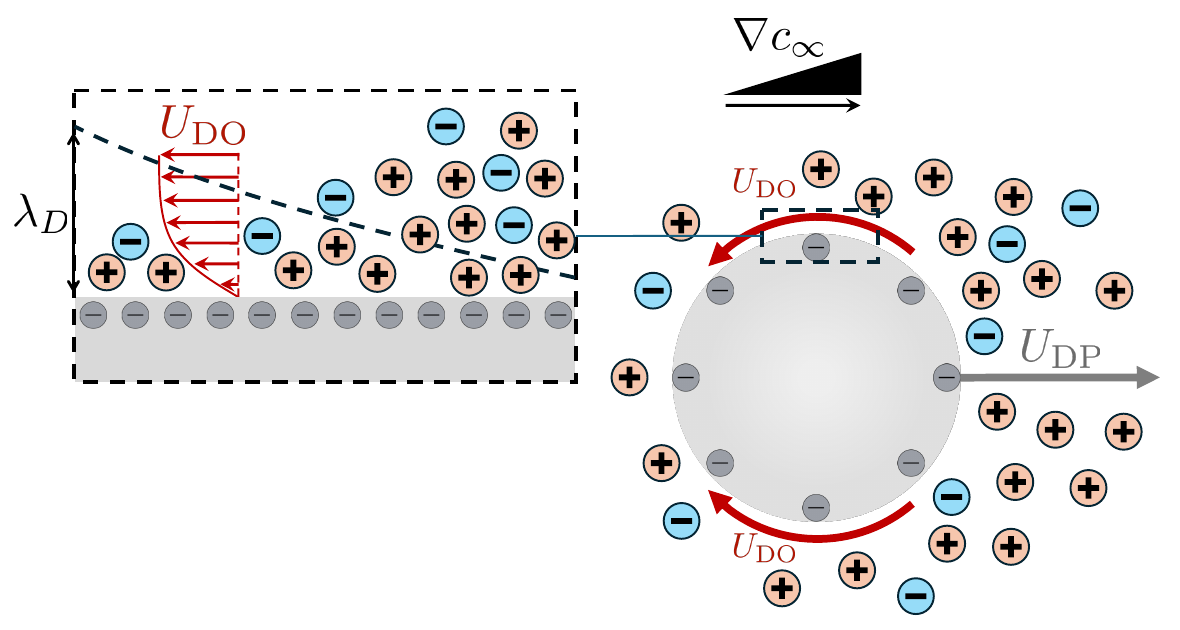}
    \caption{Bulk gradient of electrolytes drive an osmotic flow on the surface of charged surfaces. This flow consists of two contributions: a chemiosmotic flow arising from surface-solute interactions, and an electroosmotic flow due to the diffusivity difference between cations and anions. As the particle is force-free in the Stokes regime, it migrates in response to this diffusioosmotic flow in a process known as diffusiophoresis. Picture made by Haoyu Liu.}
    \label{Fig4}
\end{figure*}

Diffusiophoresis is the directed motion of a colloidal particle driven by gradients in the chemical potential or concentration of dissolved solutes. Its origin lies in a thin nanometric interfacial layer surrounding the particle. Within this layer, solutes interact with the surface electrostatically, sterically or through other short-range forces \citep{Anderson1989,Anderson82,Prieve84,Marbach19,Ault24,Shi25}. When the bulk composition varies tangentially, the interfacial excess of solute is also nonuniform and the layer drives an osmotic slip flow. Following the electrokinetic decomposition used in the colloid and soft-matter literature, the slip separates into a \emph{chemiosmotic} contribution, arising from osmotic pressure gradients inside the interfacial layer, and, for ionic solutes, an \emph{electroosmotic} contribution, arising from the diffusion potential set up when anions and cations diffuse at different rates (Fig.~\ref{Fig4})\citep{Squires16,Marbach19,Shi25}. A freely suspended particle translates as the force-free counterpart of these slip flows; a stationary wall subjected to the same tangential gradients drives \emph{diffusioosmosis}. Thus diffusiophoresis and diffusioosmosis are the particle and wall manifestations of the same interfacial transport mechanism, and both must be considered in confined geometries \citep{Marbach19,Shim22,Ault24}.

For dilute binary electrolytes in the thin-Debye-layer, low-solute-P\'eclet-number limit, the phoretic drift velocity of a particle takes the compact ``log-sensing'' form
\begin{equation}
  \mathbf{u}_{\mathrm{dp}} \;=\; \Gamma_p \, \nabla \ln c \;=\; \frac{\Gamma_p}{c}\,\nabla c ,
  \label{eq:logsensing}
\end{equation}
where $c$ is the local electrolyte concentration and $\Gamma_p$ is the diffusiophoretic mobility, which carries units of a diffusivity ($\mu\mathrm{m}^2/\mathrm{s}$) and is typically of order $10^{-10}$--$10^{-9}~\mathrm{m}^2/\mathrm{s}$ \citep{Abecassis08,Velegol16}. The logarithmic dependence is the single most consequential feature of Eq.~\eqref{eq:logsensing}: the particle responds to the \emph{relative} change in concentration, so that a tenfold contrast produces the same drift whether the concentrations are micromolar or millimolar, and the forcing remains appreciable even where the absolute concentration, and hence the linear gradient, has become small. It is this property that keeps phoresis alive along the diffuse, low-amplitude fronts that porous-media mixing produces. We note, however, that the diffusiophoretic mobility is also concentration-dependent and tends to zero as solute concentration vanishes, avoiding unphysical divergence \citep{Anderson82,Prieve84,Gupta20}. Further, this expression assumes a thin interfacial layer compared with the particle size and a phoretic motion slow enough that the solute distribution around the particle is quasi-steady; when advection competes with diffusion at the particle scale, finite-P\'eclet corrections modify both the magnitude and, in strong gradients, the character of the response \citep{Khair13,Michelin14}.

For a symmetric $Z\!:\!Z$ electrolyte the mobility comprises two physically distinct contributions, $\Gamma_p = \Gamma_{\mathrm{chemi}} + \Gamma_{\mathrm{electro}}$. The chemiphoretic term arises from the osmotic imbalance of ions within the interfacial layer; the electrophoretic term arises because the cation and anion generally diffuse at different rates, setting up a spontaneous electric field, i.e., the diffusion potential, that drags the charged particle along. For a thin double layer, both contributions are captured by the classical Prieve--Anderson expression \citep{Prieve84,Anderson1989,Velegol16}
\begin{equation}
  \Gamma_p=\frac{\varepsilon}{\mu}\left(\frac{k_{B}T}{Z e}\right)^{2}
  \Bigg[\,\underbrace{\beta_{\mathrm{ion}}\,\frac{Z e\zeta}{k_{B}T}}_{\text{electrophoretic}}
  \;+\;\underbrace{4\ln\cosh\!\left(\frac{Z e\zeta}{4k_{B}T}\right)}_{\text{chemiphoretic}}\Bigg],
  \label{eq:fullmobility}
\end{equation}
where $\varepsilon$ is the permittivity, $\mu$ the viscosity, $\zeta$ the particle zeta potential and $\beta_{\mathrm{ion}}=(D_+-D_-)/(D_++D_-)$ the cation--anion diffusivity contrast. The chemiphoretic term $4\ln\cosh(Z e\zeta/4k_BT)\ge0$ is non-negative and therefore drives the particle up the concentration gradient for either sign of $\zeta$ in this ideal electrolyte model. The electrophoretic term can be positive or negative depending on ion diffusivities and particle charge. Their sum sets the sign of $\Gamma_p$, so electrolyte diffusiophoresis can reverse direction as the salt, valence, pH, zeta potential or background concentration changes \citep{Anderson1989,Velegol16,Shim22b,Lee23,Shi25}. For non-electrolytes there is no universal sign: adsorption or attraction and depletion or repulsion give opposite excess-solute profiles in the interfacial layer and can move particles either up or down the solute chemical-potential gradient \citep{Anderson82,Marbach19,Sear17,Williams20}. Throughout this chapter we describe a particle as experiencing \emph{attractive} diffusiophoresis when it migrates toward the imposed high-concentration side of a front and \emph{repulsive} diffusiophoresis when it migrates away from it; this phenomenological terminology refers to the sign of $\Gamma_p$ in the imposed gradient, not necessarily to a microscopic attractive or repulsive interaction.

The solute diffusivity $D_s$ governs how rapidly a concentration gradient relaxes and disperses, while the particle diffusivity $D_p$ governs the colloid's own random spreading. Because colloids are far larger than simple ions, $D_p \ll D_s$, commonly by three to four orders of magnitude, so that a solute gradient can carry a particle over distances it cannot reach by Brownian motion in finite time. This separation is quantified by the two diffusion times across a pore or channel of size $L$, $\tau_s = L^2/D_s$ and $\tau_p = L^2/D_p$, with $\tau_s \ll \tau_p$. The strength of the phoretic response itself is set by the dimensionless ratio $\Gamma_p/D_s$, which recurs as the controlling exponent in the quantitative results for both dead-end-pore clearance and macroscopic dispersion. A micron-scale polystyrene colloid in a lithium chloride gradient has $\Gamma_p$ of order $0.6 \times 10^{-9}$~m$^2$/s, while the ambipolar salt diffusivity of LiCl is $D_s \approx 1.4\times10^{-9}$~m$^2$/s, so that $\Gamma_p/D_s$ is typically of order $0.1$--$0.5$ \citep{Shin16,Gupta20,Li26,Alipour26}. We use $c_0$ and $c_1$ for the two characteristic concentrations bounding a problem, i.e., the resident and displacing solutions, and write their ratio as $\beta \equiv c_1/c_0$. The most useful control parameters can be summarized as a regime map (Table~\ref{tab:regime_map}).

\begin{table*}[t]
\centering
\caption{Regime map for porous-media diffusiophoresis. The quantities listed here are the minimum set needed to interpret the experiments and scalings discussed in the rest of the chapter.}
\label{tab:regime_map}
\small
\renewcommand{\arraystretch}{1.15}
\begin{tabular*}{\textwidth}{@{\extracolsep{\fill}}p{0.22\textwidth}p{0.28\textwidth}p{0.42\textwidth}@{}}
\hline
\textbf{Quantity} & \textbf{Physical meaning} & \textbf{What it controls} \\
\hline
$\Gamma_p/D_s$ & phoretic strength relative to solute diffusion & cumulative displacement, dead-end clearance and dispersion modification \\
$c_1/c_0$ & imposed concentration contrast & sign and magnitude of forcing; enters through logarithmic ratios \\
$\mathrm{Pe}_s = U\ell/D_s$ & solute advection relative to diffusion & persistence, stretching and dispersal of solute fronts \\
$\tau_s=L^2/D_s$ and $\tau_p=L^2/D_p$ & solute and particle diffusion times & separation between rapid gradient relaxation and slow Brownian escape \\
$T/\tau_s$ & front-passage time relative to solute diffusion in a stagnant pore & sharp-front versus diffuse-front clearance regimes \\
$\alpha$ & fraction of stagnant or dead-end pore volume & relative importance of mobile--immobile exchange \\
$|\Gamma_w|/|\Gamma_p|$ & wall diffusioosmotic mobility relative to particle mobility & focusing, reversal and wall-dominated transport in tight pores \\
$a/d$ and $\lambda_D/a$ & particle size and interfacial-layer thickness relative to pore and particle scales & confinement, filtration, thin-layer validity and hydrodynamic hindrance \\
\hline
\end{tabular*}
\end{table*}

\section[Microfluidic platforms]{Microfluidic platforms and measurement of diffusiophoretic transport}\label{sec:methods}

Microfluidics is well suited for studying diffusiophoretic transport of colloids in porous media as the geometric features of the medium can be prescribed so velocity fields and solute profiles can be computed and compared directly with measurement; further, the relevant length scales (microns to millimeters) and times (seconds to hours) are experimentally accessible. This section describes the fabrication, the design of ordered-to-disordered geometries, the imposition of sharp versus diffuse fronts, imaging, and the extraction of transport statistics together with the control protocol that makes it possible to separate the competing mechanisms discussed later.

\subsection{Fabrication of microfluidic porous models}\label{sec:fab}
\begin{figure*}[htb!]
    \centering
    \includegraphics[width=1\textwidth]{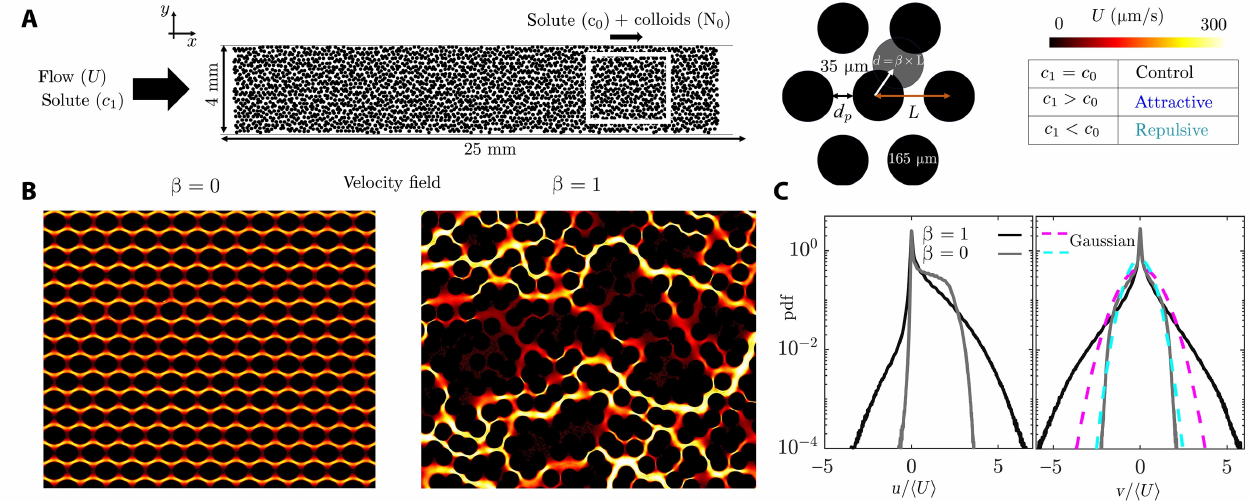}
    \caption{(a) Representative microfluidic platform for studying diffusiophoretic transport in porous media. (b) Flow velocity patterns and the corresponding distribution (c) in ordered and disordered lattices. Panels adapted from \citet{Alipour26} with permission.}
    \label{Fig5}
\end{figure*}

The porous models, or ``micromodels,'' are quasi-two-dimensional channels patterned with arrays of solid posts (Fig.~\ref{Fig5}). They are made by standard soft lithography: a master bearing the post pattern is produced by photolithography in a photoresist on a silicon wafer, poly(dimethylsiloxane) (PDMS) is cast against the master and cured, and the molded PDMS slab is bonded to a glass coverslip after oxygen-plasma activation \citep{Xia98}. The result is a sealed network of uniform depth, typically tens of micrometers, with in-plane post diameters and throat widths of tens to hundreds of micrometers. In the experiments reviewed below, the obstacles are circular posts of diameter $2R\approx165~\mu$m arranged with a mean pore (throat) size $d_p\approx35~\mu$m in a channel of depth $\approx50~\mu$m, and the displacing solution is driven at a fixed volumetric rate (of order $10^{-2}~\mu$L/s with a syringe pump), so that the pore-scale Reynolds number is $\ll1$ and the flow is everywhere in the Stokes regime~\citep{Alipour26}. PDMS is optically transparent and gas permeable, and both PDMS and glass present native negative surface charge in contact with aqueous electrolytes. The shallow, planar geometry makes the entire pore space accessible to an inverted microscope in a single focal plane, at the cost of imposing two-dimensionality, whose consequences for mixing and migration are discussed in Section~\ref{sec:outlook}.

\subsection{Designing ordered-to-disordered pore structures}\label{sec:disorder}
A central experimental advantage of using micromodels is that the microstructure can be tuned, from perfectly ordered to strongly disordered, allowing the role of geometric heterogeneity to be isolated. To introduce disorder, we begin with an ordered lattice, e.g., a hexagonal array of circular posts, displacing each post from its lattice site by a random vector whose amplitude is set by a single control parameter, i.e., each post is displaced by a random vector of magnitude up to $\chi L$, where $L=2R+d_p$ is the unperturbed lattice spacing and $0\le\chi\le1$, where $d_p$ is the throat width. Introducing disorder, broadens the throat-width distribution, leading to the emergence of preferential paths and slow zones, and beyond a threshold the perturbed lattice begins to occlude throats and create dead-end pores (Fig.~\ref{Fig5})~\citep{Alipour26}. Because $\chi$ controls the breadth of the local-velocity distribution, sweeping it maps the velocity heterogeneity of the medium onto a single, reproducible axis. This design is what later permits the decisive comparison between geometries that do and do not contain dead-end pores while holding the rest of the flow statistics broadly comparable.

\subsection{Imposing solute gradients: sharp versus diffuse fronts}\label{sec:fronts}
Diffusiophoresis is driven by the solute gradient, so controlling the steepness of that gradient is the core of the experiment. A background solution of concentration $c_0$ initially saturates the device; a displacing solution of concentration $c_1$ is then introduced at a constant volumetric flow rate, establishing a front that propagates through the network. The contrast $\beta=c_1/c_0$ sets the magnitude of the forcing and its sign convention (attractive for $c_1>c_0$, repulsive for $c_1<c_0$). The steepness of the front is the second, independent knob. A sharp front is produced by switching rapidly to the displacing solution close to the test section, so that the colloids see a large but short-lived gradient. A diffuse front is produced by letting the solute disperse before it reaches the colloids: passing the displacing stream through a long upstream channel, i.e., a ``dispersion tube'', allows Taylor dispersion to broaden the front in a controlled way, so that the colloids experience a weaker gradient sustained over a much longer transition time $T$. This pairing of sharp and diffuse fronts at fixed $\beta$ is precisely what allows the cumulative, time-integrated forcing to be probed, and it underlies the dead-end-pore results of Section~\ref{sec:deadend}. We note that the heterogeneous velocity field of porous medium further enhances the longitudinal dispersion of the solute front, leading to its broadening in our porous media experiments. For experiments that require the solute field itself to be visualized, the electrolyte is co-injected with a fluorescent tracer of known diffusivity, so that the evolving concentration profile can be imaged alongside the particles.

\subsection{Colloids, solutes, and fluorescence imaging}\label{sec:imaging}
The model colloids are typically fluorescent polystyrene spheres, from tens of nanometers to a few micrometers in diameter, whose surface (zeta) potential is characterized independently as it sets both the sign and the magnitude of $\Gamma_p$. A common choice is carboxylate-functionalized polystyrene spheres of diameter $\approx1~\mu$m (with excitation/emission near $540/560$~nm), suspended at a volume fraction of order $0.1\%$ so that the colloids behave as non-interacting point particles: the diffusiophoretic flow disturbance around a particle decays as $1/r^{3}$, much faster than the $1/r$ of a body force, so particle--particle and particle--wall hydrodynamic couplings are negligible at this dilution \citep{Shin16}. The solutes are simple salts (for example LiCl, NaCl, KCl) chosen for their differing ionic-diffusivity contrasts, which tune the electrophoretic contribution to the mobility. Typical electrolyte concentrations span $0.01$--$100$~mM, a range over which double-layer screening prevents salt-induced aggregation; and because polystyrene ($\approx1050~\mathrm{kg/m^3}$) is only slightly denser than water, the Stokes settling speed is $\mathcal{O}(10)~$nm/s, so sedimentation is negligible over the $10$--$20$~min duration of a typical experiment. Imaging is by epifluorescence on an inverted microscope. Where particle and solute must be distinguished, spectrally separated dyes are used and imaged in alternation. 

\subsection{Extracting transport statistics: breakthrough curves and dispersion}\label{sec:btc}
\begin{figure*}[htb!]
    \centering
    \includegraphics[width=1\textwidth]{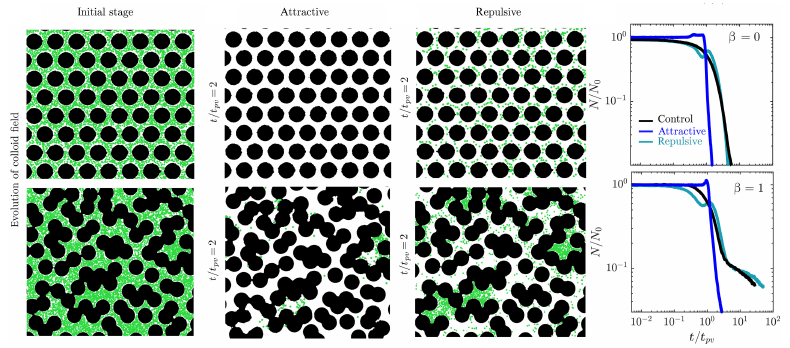}
    \caption{Diffusiophoretic transport of colloids in ordered/disordered micromodels. Green dots represent the colloidal particles. The evolution of total number of particles $N(t)$ normalized by the initial number $N_0$ shows a strong influence of solute gradients; colloids are removed much faster in the attractive case ($c_1/c_0>1$) than the repulsive ($c_1/c_0<1$) and control ($c_1/c_0=1$) cases. Panels adapted from \citet{Alipour26} with permission.}
    \label{Fig6}
\end{figure*}

Two complementary observables summarize colloid transport through the medium. At the pore scale, particle tracking yields individual trajectories, from which one obtains velocity distributions, and residence times. At the macroscale, the natural observable is the breakthrough curve (BTC): the normalized concentration of colloids arriving at the outlet, $N/N_0$, plotted against time made dimensionless by the pore-volume time $t_{\mathrm{pv}}$ (the pore volume of the device divided by the volumetric flow rate), so that one dimensionless time unit corresponds to one full flushing of the pore space (Fig.~\ref{Fig6}). The shape of the BTC encodes the transport: a steep early decay reflects fast preferential pathways, while a long tail reflects particles detained in slow zones and dead-end pores, i.e., the signature of non-Fickian transport. The macroscopic (hydrodynamic) dispersion coefficient $D^{*}$ is obtained by fitting the BTC with the 1D advection-diffusion model, and it is the quantity whose dependence on the solute gradient is reported in Section~\ref{sec:flow}. Comparing a gradient-driven experiment with a control experiment performed at uniform concentration ($\beta=1$) isolates the diffusiophoretic contribution from the purely hydrodynamic one (Fig.~\ref{Fig6}).

\section[Diffusiophoresis in dead-end pores]{Diffusiophoresis in dead-end pores exposed to diffuse fronts}\label{sec:deadend}

\begin{figure*}[htb!]
    \centering
    \includegraphics[width=1\textwidth]{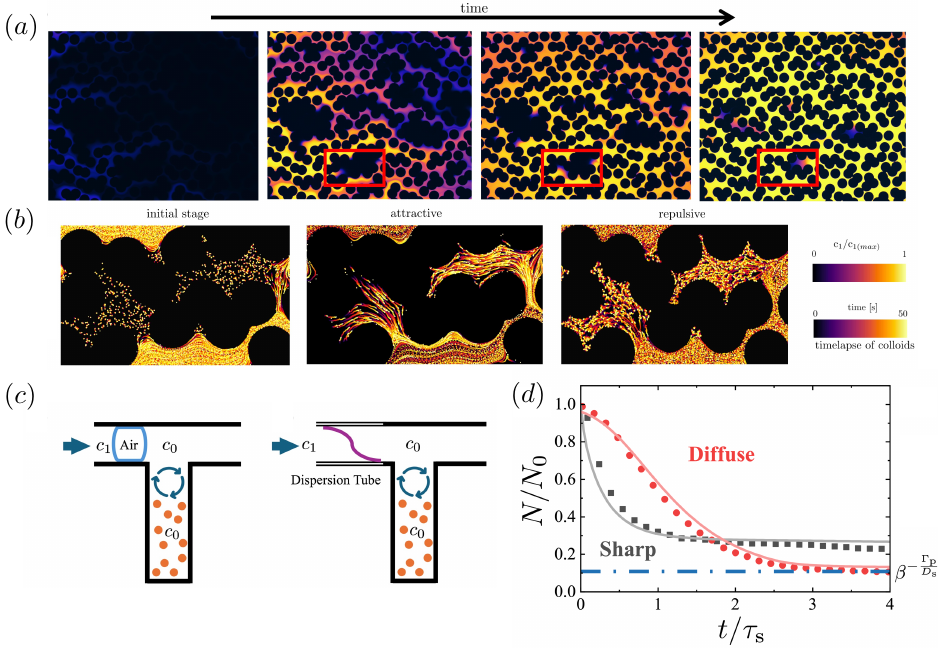}
    \caption{(a) Flow velocity disorder and heterogeneous mixing enhances the dispersion of solute fronts in porous media, weakening the spatial gradients, but making them more persistent in time. (b) These gradients remain "strong enough" to drive diffusiophoretic migration of colloids in dead-end pores within the medium. Comparing sharp versus diffuse solute fronts (c), we found that while sharp fronts are more efficient in particle removal over short times ($t<\tau_s$), diffuse fronts are more efficient over long times ($t>\tau_s$), where the residual fraction of particles asymptotes to $\beta^{-\Gamma_p/D_s}$ with $\beta = c_1/c_0$. Panels adapted from \citet{Alipour26,Li26} with permission.}
    \label{Fig7}
\end{figure*}

Dead-end channels and pores offer a canonical platform to study diffusiophoresis and diffusioosmosis. Many microfluidic experiments have used sharp solute fronts to probe the migration of colloids and emulsions in response to the resulting gradients \citep{Shin16,Kar15,Battat19,Ault17}. In porous media, however, flow disorder and mechanical dispersion typically produce diffuse rather than sharp fronts, weakening the instantaneous concentration gradients (Fig.~\ref{Fig7}(a,b)). While experiments and simulations have demonstrated diffusiophoretic migration of colloids from dead-end pores \citep{Park21,Jotkar24,Jotkar24b}, a central question remained unanswered: does dispersion weaken diffusiophoretic extraction from dead-end pores, rendering the effect negligible at larger scales?

We recently addressed this question in an idealized dead-end-pore geometry using analytical modeling, numerical simulations, and microfluidic experiments \citep{Li26}; we note that related work has examined the influence of time-dependent solute profiles on particle dynamics inside dead-end pores \citep{Migacz24}. The key insight is that the relevant quantity is not the instantaneous gradient but the cumulative, time-integrated phoretic forcing. A sharp front produces a large gradient but only for the short time over which solute diffuses along the pore, $\tau_s = L^2/D_s$. A diffuse front produces weaker gradients but sustains them over a longer transition time $T$ (Fig.~\ref{Fig7}(c)). When $T \gg \tau_s$, the cumulative drift can be larger, and removal more effective, even though the phoretic velocity is smaller at every instant. Sharp gradients thus win at short times ($t<\tau_s$), but diffuse gradients can remove more particles at long times ($t>\tau_s$) (Fig.~\ref{Fig7}(d)).

A useful way to build intuition is to follow particles initially near the closed end of the pore. Under a sharp front, particles feel a strong initial pull toward the mouth, but the gradient collapses as the solute front relaxes and as the particles move into regions of weaker forcing, leaving a residual population stranded near the entrance. Under a diffuse front, the gradient along the pore stays approximately linear for a long time, driving a weak but sustained drift that produces more uniform clearance. The contrast is a direct manifestation of the log-sensing law, Eq.~\eqref{eq:logsensing}: because the drift depends on $\nabla\ln c$ rather than $\nabla c$, a front that is weak in absolute terms but broad in time can integrate to a larger net displacement than a strong but fleeting one.

The long-time efficiency of this process can be captured in closed form \citep{Li26}. Consider a one-dimensional dead-end pore of length $L$, with coordinate $x$ running from the closed end ($x=0$) to the mouth ($x=L$), and a non-diffusive particle that moves with the local phoretic velocity $u_{\mathrm{dp}} = \Gamma_p\,\partial_x \ln c$. The solute obeys diffusion within the pore, and for a front that varies slowly compared with $\tau_s$ the solute profile is quasi-steady, evolving through a family of self-similar shapes as the mouth value rises from $c_0$ toward $c_1$. In this quasi-static limit the particle trajectory is obtained by integrating $\mathrm{d}x/\mathrm{d}t = \Gamma_p\,\partial_x \ln c$, so that the mapping from a particle's initial position to its final position is governed entirely by ratios of the local concentration. The fraction of particles ``cleared'' is the fraction that reaches the mouth before the gradient vanishes. The long-time upper bound on the clearance efficiency of non-diffusive particles is
\begin{equation}
\mathcal{E}_{\mathrm{cl}} \;=\; 1 - \left(\frac{c_1}{c_0}\right)^{-\Gamma_p/D_s},
\label{eq:clearance}
\end{equation}
where $\mathcal{E}_{\mathrm{cl}}$ is the clearance efficiency, $c_0$ and $c_1$ are the initial and displacing concentrations, $\Gamma_p$ is the diffusiophoretic mobility, and $D_s$ is the solute diffusivity \citep{Li26}. Equivalently, the residual fraction left behind is $N_1/N_0 = (c_1/c_0)^{-\Gamma_p/D_s} = \beta^{-\Gamma_p/D_s}$, with $N_0$ the initial and $N_1$ the residual particle number in the pore. Taking the representative value $\Gamma_p/D_s\approx 0.2$ with $\beta=100$ leaves a residual fraction $\beta^{-\Gamma_p/D_s}=100^{-0.2}\approx0.4$: roughly $60\%$ of the trapped colloids are flushed from the pore. The single dimensionless group $\Gamma_p/D_s$, i.e., the ratio of phoretic strength to solute diffusivity, controls the outcome, and the dependence on concentration enters through the ratio $\beta$. At still longer times ($t>\tau_p$), particle diffusion clears the residual population and sets the ultimate approach to complete removal. \textbf{This answers the first guiding question of the chapter: dispersion does not defeat diffusiophoretic extraction from dead-end pores}. By trading gradient amplitude for duration, diffuse fronts can be \emph{more} effective than sharp ones at the long times that matter in porous media.

\section[Cross-streamline migration]{Diffusiophoresis in porous media flows: cross-streamline migration and macroscopic dispersion}\label{sec:flow}

In macroscopic transport through porous media, travel times are controlled by how particles sample the velocity heterogeneity of the flowing backbone. A natural intuition holds that diffusiophoresis should be negligible there, because mean flow speeds in microfluidic porous media and preferential pathways are often $\mathcal{O}(10^2$--$10^3)~\mu\mathrm{m}/\mathrm{s}$, whereas diffusiophoretic speeds estimated from calibrated salt gradients are commonly $\mathcal{O}(1$--$10)~\mu\mathrm{m}/\mathrm{s}$ \citep{Shin16,Ault24,Alipour26}. Our experiments and simulations show that this intuition is wrong \citep{Alipour26}: small transverse drifts of colloids in response to solute gradients re-partition particles among the slow and fast streamlines of the preferential pathways, and this rearrangement has outsized consequences. Macroscopic dispersion of colloids can then deviate by orders of magnitude from the predictions of classical models that ignore diffusiophoresis.

We probed colloid transport in microfluidic channels patterned with ordered/disordered arrays of obstacles. Beginning from ordered hexagonal lattices, we gradually perturbed the obstacle positions by an amplitude $\chi$, the disorder strength, to create disordered lattices, and imposed solute gradients by displacing a background solution of concentration $c_0$ with one of concentration $c_1$ at constant flow rate.  Measuring colloid transport statistics across a range of disorder strengths and concentration ratios then reveals how the interplay of diffusiophoresis and flow disorder shapes transport.

\begin{figure*}[htb!]
    \centering
    \includegraphics[width=1\textwidth]{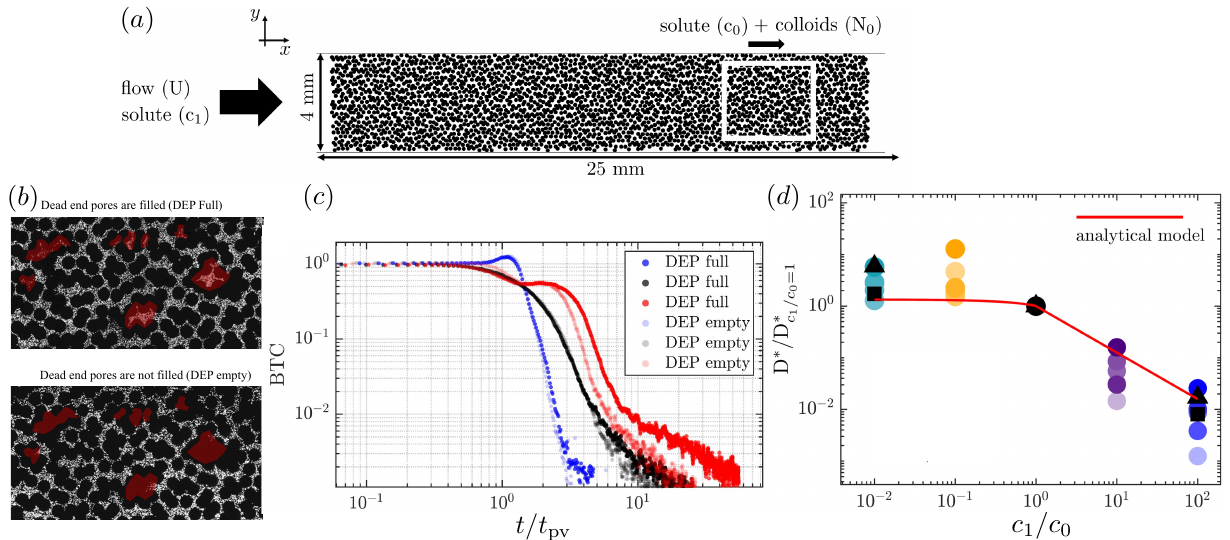}
    \caption{(a) We study the influence of solute gradients on the transport of colloids in microfluidic channels embedded with ordered/disordered lattice of obstacles. (b) To isolate the role of dead-end pores, we performed identical experiments with dead-end pores initially filled/empty of colloids. (c) The influence of solute gradients on the travel time of colloids as characterized by the breakthrough curves is insensitive to whether dead-end pores are filled/empty, suggesting that cross-streamline migration of colloids within the preferential flow pathways is the key mechanism for the macroscopic impact of diffusiophoresis. (d) This diffusiophoretic cross-streamline migration mechanism leads to changes of the macroscopic dispersion by orders of magnitude compared to the control case without solute gradients. Panels adapted from \citet{Alipour26} with permission.}
    \label{Fig8}
\end{figure*}

In the disordered lattices, where dead-end pores are present, the fraction of colloids in the stagnant zones is indeed modulated by solute gradients. In the attractive case ($c_1>c_0$), phoretic migration of colloids out of dead-end pores can even suppress the non-Fickian tail of the breakthrough curve, as particles escape the traps riding the solute wave; the residual fraction in the dead-end pores scales as $N_1/N_0 \sim \alpha\,(c_1/c_0)^{-\Gamma_p/D_s}$, where $\alpha$ is the fraction of dead-end pores and $N_0$, $N_1$ are the initial and post-equilibration particle numbers in them (Fig.~\ref{Fig7}). Surprisingly, however, we also observed a strong footprint of solute gradients on transport in ordered and weakly disordered lattices, where no dead-end pores exist, raising the question of whether the influence of diffusiophoresis is confined to dead-end pores, as is commonly assumed.

To settle this, we repeated the experiments on disordered lattices but left the dead-end pores empty of colloids (Fig.~\ref{Fig8}(a,b)). If diffusiophoresis acted only through dead-end pores, leaving them empty should erase any effect of the solute gradient on transport. Instead, we observed the same change in colloid travel time, as characterized by the breakthrough curves, as when the dead-end pores were initially filled (Fig.~\ref{Fig8}(b,c)). The strong influence of solute gradients on transport through the medium is therefore not due to phoretic exchange in and out of dead-end pores. This unexpected coincidence of the filled and empty curves poses the central question: how does the influence of diffusiophoresis extend beyond dead-end pores?

The answer is found by examining colloid transport in a single channel flow \citep{Alipour26, Li26b}, which stands in for a preferential pathway in the network. Earlier studies treated this geometry either at very early times, shorter than the solute diffusion time across the channel ($t<\tau_s$) \citep{Migacz22}, or in the asymptotic Taylor-dispersion regime ($t\gg\tau_s$) under the assumption of a uniform solute field  across the channel \citep{Chu21,Chu22}. A solute front in confined pressure-driven flow, however, experiences both shear and diffusion, and even after the solute has homogenized across the channel width, weak transverse gradients persist wherever a longitudinal gradient remains. As particles advect downstream they therefore feel a sustained transverse solute gradient that drives cross-streamline diffusiophoretic migration.

The consequence for macroscopic dispersion can be captured by a simplified model \citep{Alipour26}. In pressure-driven flow in a channel the streamwise velocity varies parabolically across the gap, so a particle's average speed, and hence its contribution to dispersion depends on which streamlines it samples. A transverse phoretic drift $v_{\mathrm{dp}} = \Gamma_p\,\partial_y \ln c$ systematically moves particles across these streamlines: in the attractive case it focuses them, reducing the spread of velocities the population samples and therefore reducing dispersion, while in the repulsive case it defocuses them, weakly increasing the spread. For the attractive case, the macroscopic dispersion is reduced as
\begin{equation}
D^{*}/D^{*}_c \;\sim\; \left(\frac{c_1}{c_0}\right)^{-\Gamma_p/D_s},
\label{eq:disp-att}
\end{equation}
and, for a repulsive front, a weak enhancement,
\begin{equation}
D^{*}/D^{*}_c \;\sim\; \left( 7\sqrt{3} - 3\left(\frac{c_1}{c_0}\right)^{\Gamma_p/D_s} \right)\big/9 ,
\label{eq:disp-rep}
\end{equation}
where $D^{*}_c$ is the dispersion coefficient of the control without solute gradients. The same dimensionless group $\Gamma_p/D_s$ and the same concentration ratio $\beta$ that governed dead-end-pore clearance reappear here, now controlling a macroscopic transport coefficient. Numerically, the representative value $\Gamma_p/D_s\approx0.2$ with $\beta=100$ already reduces the attractive-front dispersion to $D^{*}/D^{*}_c\approx 0.4$ (Eq.~\eqref{eq:disp-att}); when the phoretic response is stronger, $\Gamma_p/D_s\to\mathcal{O}(1)$, the same contrast drives $D^{*}/D^{*}_c$ toward $10^{-2}$, the order-of-magnitude suppression of dispersion measured in the experiments. This simple channel model agrees qualitatively with the measured changes in breakthrough and velocity statistics (Fig.~\ref{Fig8}(d)), confirming that solute gradients can strongly reshape macroscopic transport relative to the gradient-free control. While the channel analogue does not capture full network-level mechanical dispersion \citep{Saffman59,Koch85,Dentz18}, it provides a transparent bridge from pore-scale cross-streamline migration to a measurable macroscopic coefficient, and the mechanism it isolates is active within the preferential pathways that resemble channels embedded in the network.

Cross-streamline migration thus changes which streamlines, and which velocities, particles sample over time. In the attractive case ($c_1>c_0$) particles drift toward faster streamlines, raise their mean speed, and exit the medium sooner; in the repulsive case the trend reverses. Because travel time depends on the cumulative velocity history, a modest systematic transverse drift produces large macroscopic effects, including pronounced changes in early-time breakthrough that persist for both ordered and disordered lattices. Solute gradients therefore lead to two outcomes: they change the macroscopic dispersion of colloids by orders of magnitude, and they suppress the influence of geometric disorder on that dispersion. Together, these results imply that traditional models of colloid transport in porous media need to be revisited when they neglect diffusiophoresis and describe transport mainly through a velocity distribution set by geometric disorder. \textbf{This answers the second guiding question of the chapter: the influence of diffusiophoresis is emphatically not limited to dead-end pores}.\\

\paragraph{Practical considerations and challenges.}

Quantifying diffusiophoresis in porous media is complicated by the fact that several effects act at once, often with comparable magnitude. We collect here the practical issues that most frequently confound measurement and interpretation, both as guidance for experiments and as a map of where the theory could be most helpful.
\begin{itemize}
\item {Disentangling diffusiophoresis from diffusioosmosis.} Boundaries support diffusioosmotic slip, so that the imposed solute gradients drive an osmotic flow along the walls in addition to the phoretic drift of the particles. In the laboratory frame the colloid velocity is then the sum of the imposed flow, the diffusioosmotic flow, and the phoretic velocity, with $\mathbf{u}_{\mathrm{do}}$ set by the wall mobility $\Gamma_w$ and $\mathbf{u}_{\mathrm{dp}}=\Gamma_p\nabla\ln c$ by the particle mobility, so that the competition between them is governed by the signed mobilities and, for scaling purposes, by the magnitude ratio $|\Gamma_w|/|\Gamma_p|$. Near boundaries the two can either reinforce or oppose one another, and because the osmotic flow advects particles indiscriminately it can mask a genuine phoretic response. The interplay can lead to unexpected results, qualitatively changing particle trajectories, leading to the formation of vortices in microfluidic T junctions or parallel flow geometries \citep{Shin20,Migacz23}, and can even reverse the direction in which colloids are focused relative to what bulk diffusiophoresis alone would predict \citep{Liu25}. Interpreting particle motion therefore requires the wall zeta potential and channel geometry to be known, and ideally a companion simulation that includes the diffusioosmotic boundary condition \citep{Shin17b,Ault19,Shim22,Akdeniz23,Lee23,Zhang24,Chakra25}.

\item {Determining the mobility and its sign.} The mobility $\Gamma_p$ is rarely known a priori for a given particle--solute pair. As evident from Eq.~\eqref{eq:fullmobility}, $\Gamma_p$ varies with salt identity, ion valence, diffusivity contrast, background ionic strength and particle zeta potential. Because the zeta potential is itself set by surface chemistry, adsorption, pH and charge regulation, the mobility can change sign as conditions vary; a system that is attractive at one ionic strength or pH may be repulsive at another \citep{Shim22b,Lee23,Yang23,Yang24,Chakra25}. A clear example is pH: changing pH can reverse the surface (zeta) potential and hence the phoretic response \citep{Shim22b}, a mechanism that has recently been used to rationalize diffusiophoretic focusing in acid--base reaction fronts \citep{Shi16,Coleman25}. Independent calibration, for example tracking band displacement or focusing in a known one-dimensional gradient, is therefore essential, and the calibration must span the concentration range of the desired experiment, since assuming a constant $\Gamma_p$ across a large contrast $\beta$ can introduce error \citep{Akdeniz23,Lee23}.
\end{itemize}

\section{Diffusiophoretic transport of emulsions}\label{sec:emulsions}

The colloids considered so far are typically rigid and chemically passive. Emulsion droplets, i.e., oil drops dispersed in water, break both assumptions, and in doing so open a frontier that is both rich in physics and directly relevant to enhanced oil recovery, remediation and drug delivery. Two features distinguish an emulsion droplet from a rigid colloid in a solute gradient. First, its interface is mobile: rather than a no-slip condition, the boundary enforces continuity of velocity and tangential stress, so interfacial diffusioosmotic stresses can drive internal circulation, and the resulting drift depends on the viscosity ratio between the drop and the surrounding fluid as well as on the solute gradient and solute--interface interaction \citep{Yang18,Khair13,McKenzie22,Marbach20,Shi25}. Second, the same solute gradient can produce a surface-tension gradient along the interface and hence a Marangoni stress \citep{Lohse20,Duong26}. These two drivers can have a common chemical cause, but they are distinct: diffusioosmotic slip originates in osmotic stresses within an interfacial layer, whereas Marangoni flow originates in the dependence of interfacial tension on local interfacial composition, commonly through adsorption and desorption of solutes or surfactants. In emulsions inside pores, these stresses couple to coalescence, dissolution, aggregation and trapping \citep{Kar15,Park21,Duong26}.

\begin{figure*}[htb!]
    \centering
    \includegraphics[width=1\textwidth]{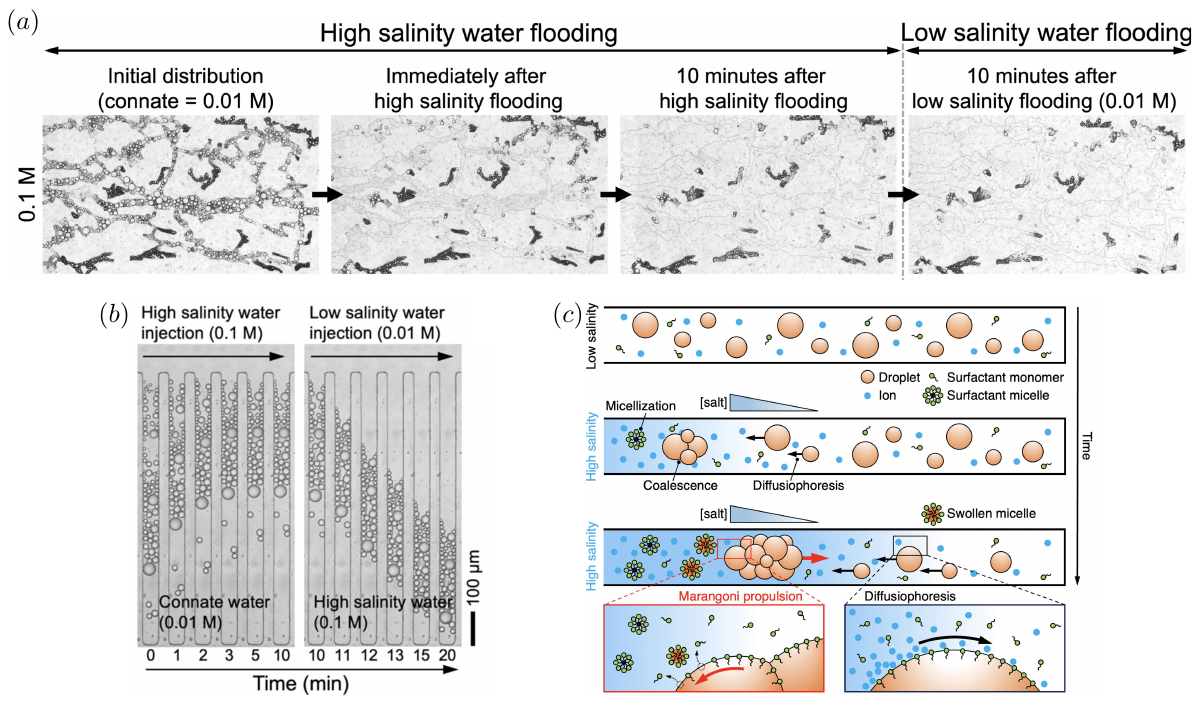}
    \caption{Droplets and emulsions in solute gradients. (a,b) Salinity fronts can mobilize or aggregate oil drops in porous micromodels and dead-end pores. (c) Surfactant-laden drops can show competition between diffusiophoresis and Marangoni propulsion. Panels adapted from \citet{Park21,Duong26} with permission.}
    \label{Fig9}
\end{figure*}

For a charged emulsion droplet in an electrolyte gradient, combined theory and experiment show that the diffusiophoretic response differs from that of an otherwise equivalent rigid particle because electric and viscous stresses drive flow both outside and inside the droplet \citep{Yang18}. The speed depends on viscosity ratio, and in the large-viscosity-ratio limit the result approaches the rigid-particle value. For the charged drops studied by \citet{Yang18}, electrokinetic diffusiophoresis dominated the competing Marangoni contribution, but which mechanism prevails is not universal; a crossover is expected as surface charge or surfactant coverage vary. For a deformable droplet in a neutral-solute gradient, a uniform gradient translates the drop without deforming it at leading order, whereas a nonuniform gradient can deform the interface, with the sign and magnitude set by solute--drop interactions and capillarity \citep{McKenzie22}.

Surfactants make the problem even richer. Their behavior at interfaces couples adsorption, desorption, micellization and surface rheology, all of which control how droplets respond to a gradient \citep{Manikantan2020}. Ionic surfactants can contribute to diffusiophoresis through electrolyte-like mechanisms, but salt concentration also sets the critical micelle concentration and the equilibrium surfactant coverage of the interface, so a salinity gradient can indirectly modulate Marangoni driving \citep{Shi21,Duong26}. Recent experiments on surfactant-laden oil droplets under salinity gradients show that diffusiophoresis and Marangoni propulsion can compete: at high salinity, salting-out-induced changes in micelles and interfacial surfactant coverage can trigger aggregation, nonmonotonic motion and even reversal (Fig.~\ref{Fig9} (c)) \citep{Duong26}. Related surfactant-gradient studies of colloids show that adsorption history and polymer--surfactant complexation can also change the direction and magnitude of particle motion \citep{Nery17,Yang23,Yang24}. This is directly relevant to porous media because salinity gradients and surfactants are common during chemical flooding.

The existing porous-media evidence for emulsion droplets remains limited but suggestive. Microfluidic studies of salinity-induced oil recovery showed that gradients created during high- and low-salinity flooding can mobilize oil droplets trapped in dead-end regions, while high salinity can also promote aggregation and pore blockage (Fig.~\ref{Fig9} (a,b)) \citep{Park21}. The emulsion problem therefore illustrates the dual role of chemistry: gradients can drive transport, while concentration itself changes emulsion stability and interfacial properties. A full account of oil transport in porous media must therefore include phoretic migration, Marangoni stresses, coalescence, aggregation, wetting transitions, capillarity and hydrodynamics.

\section[confined and living media]{Diffusiophoresis across confined and living media}\label{sec:landscape}

The porous-media results above sit within a fast-growing body of work on diffusiophoresis in confined, structured and living environments. 

\begin{figure*}[htb!]
    \centering
    \includegraphics[width=1\textwidth]{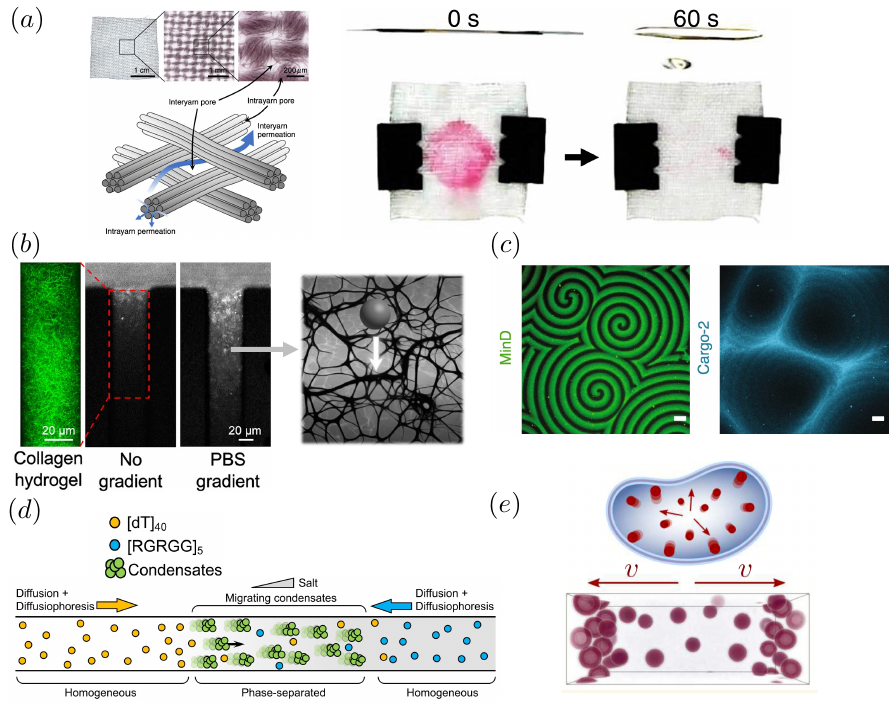}
    \caption{ Diffusiophoresis in confined and living media. (a) Surfactant rinsing helps remove stains from fabrics. (b) Colloids penetrate into a collagen hydrogel using salt gradients. (c) MinDE protein waves transport cargo on a membrane. (d) Salt gradients modulate the formation and motion of liquid condensates. (e) Reaction-driven chemical gradients could modulate transport of liquid condensates within living cells. Panels adapted from \citet{Shin18,Doan21,Ramm21,Doan24,Hafner24} with permission.}
    \label{Fig10}
\end{figure*}

\paragraph{Fabrics, cleaning and filtration.} Several of the most developed applications exploit diffusiophoresis as a means of moving particles where pressure-driven flow alone is weak. A fabric is a hierarchical porous material: most flow bypasses the intrayarn pores, leaving a stagnant core in which Brownian diffusion is too slow to explain practical cleaning times. Surfactant gradients established during rinsing provide a phoretic route for particulate removal from these low-permeability regions (Fig.~\ref{Fig10} (a)) \citep{Shin18}. This mechanism is directly analogous to dead-end-pore clearance in porous media: the solute leaves the pore space faster than the particles, and the resulting transient gradient converts fast solute diffusion into slow particle displacement.

\paragraph{Hydrogels, biofilms and soft porous matrices.} Biological porous materials introduce confinement, binding and steric obstruction. Experiments in collagen gels showed that nanoparticle diffusiophoresis can persist under tight confinement and that mobility can vary nonmonotonically with pore size or gel concentration (Fig.~\ref{Fig10} (b)) \citep{Doan21}. A complementary theoretical framework models the surrounding matrix as a Brinkman medium, predicting that hydrodynamic resistance can strongly reduce mobility relative to a free solution and that electrolyte composition can change both the magnitude and sign of transport \citep{Sambamoorthy23,Sambamoorthy25}. Biofilms provide another important example; their extracellular polymeric matrix is a crowded, heterogeneous porous medium, but imposed electrolyte gradients can nevertheless drive micro- and nanoparticles into and out of the matrix \citep{Somasundar23, Chen26}. These results suggest that diffusiophoresis may be useful for delivery, cleaning or disruption in soft biological materials, while also indicating that mobility cannot simply be imported from free-solution measurements.

\paragraph{Living and active systems.} Chemical gradients that drive passive colloids can also interact with living particles and biomolecular assemblies. Bacteria can experience diffusiophoresis through their charged surfaces, and surfactants can enhance this response by increasing cell-surface charge \citep{Doan20}. In porous habitats, motile bacteria also sense gradients through chemotaxis, so phoresis, chemotaxis and advection can coexist \citep{Anna21,Doan26}. Beyond whole cells, gradients can transport membrane-bound cargo in the MinDE system (Fig.~\ref{Fig10} (c)) \citep{Ramm21}, promote and move biomolecular condensates (Fig.~\ref{Fig10} (d,e)) \citep{Doan24,Hafner24}, and move microtubules along tubulin, RanGTP and salt gradients \citep{Shim24}. These examples show that reaction--diffusion patterning, binding, phase separation and cytoplasmic crowding can all modify the simple electrolyte picture.

The biological examples are especially relevant because they replace a rigid porous matrix by a crowded, dynamic medium. In the MinDE system, non-specific membrane cargo is redistributed by protein fluxes and can be sorted by effective size through a density-dependent friction mechanism (Fig.~\ref{Fig10} (c)) \citep{Ramm21}. For biomolecular condensates, salt or macromolecular gradients can first enrich the constituents and promote local phase separation, and can then bias the motion and lifetime of the droplets once they have formed (Fig.~\ref{Fig10} (d)) \citep{Doan24}. Reaction-driven theories extend this idea by letting fuel and waste gradients generated by biochemical cycles move liquid condensates toward or away from sources and sinks (Fig.~\ref{Fig10} (e)) \citep{Hafner24}. Microtubules provide a complementary cytoskeletal example in which gradients of tubulin, RanGTP and simple salts move extended, charged filaments at measurable speeds even in cytoplasmic extract \citep{Shim24}. Together these systems suggest that the porous-media viewpoint has an analogue in living matter: gradients are generated internally, the medium is crowded and reactive, and the transported object may change the gradient field or phase state that moves it.

\paragraph{Complex fluids and soft matrices.} Many application-relevant porous media are not Newtonian fluids in rigid pores, and it is useful to separate the distinct ways a complex medium alters phoresis. (i) \emph{Added hydrodynamic resistance.} Dissolved polymers and gel networks raise the effective viscosity and add drag; modeling the matrix as a Brinkman medium predicts that this resistance reduces the phoretic mobility relative to free solution and can change how it varies with concentration \citep{Doan21,Sambamoorthy23,Sambamoorthy25}. (ii) \emph{Modified interfacial driving force.} Polymers or macromolecules that adsorb onto, or are depleted from, the particle and pore surfaces change the solute distribution within the interfacial layer and hence the slip itself; non-adsorbing polymers, for example, drive excluded-volume diffusiophoresis down their own concentration gradient \citep{Sear17,Anderson82}. (iii) \emph{Coupled solute and microstructure.} Because salts and polymers interact, an imposed salt gradient can also be a gradient in polymer conformation and local rheology; in biofilms or extracellular matrices the matrix can additionally store elastic stress and bind particles \citep{Katke24,Chen26}. Distinguishing whether the matrix merely hinders motion, so that an effective mobility may suffice, or actively reshapes the driving force is what determines whether gel and biofilm demonstrations carry over to tissues, mucus and polymer-flooded reservoirs.

Taken together, the field has moved from a simple picture of particles in a one-dimensional salt gradient to a broader view in which chemical gradients act as transport fields in confined and heterogeneous environments. The porous-media problem combines many of the ingredients: strong velocity heterogeneity, large surface area, stagnant regions, evolving scalar gradients, particle--surface interactions and, often, reactions or multiphase interfaces.

\section{Outlook and open problems}\label{sec:outlook}
 
The results reviewed here show that diffusiophoresis can strongly influence colloid transport in porous media, and that its influence can extend well beyond the dead-end pores. They also expose the wealth of remaining open questions.

\paragraph{Algebraic versus chaotic mixing: 2D vs 3D.} The cross-streamline migration documented in Section~\ref{sec:flow} was established in quasi-two-dimensional micromodels, and, as emphasized throughout, steady two-dimensional flow can stretch material lines only algebraically. Steady Stokes flow through 3D porous media can be chaotic, stretching and folding fluid elements exponentially and thereby sustaining sharp microscale concentration gradients that would otherwise diffuse away \citep{Lester13,Lester16,Heyman20,Souzy:2020aa}. Since the phoretic drive is the logarithmic gradient of concentration, the transverse gradients that power cross-streamline migration may be sharper, more persistent, and organized along lamellar manifolds in three dimensions in ways no two-dimensional experiment can reveal, so the macroscopic effect could be larger, or qualitatively reorganized, and the relative weights of dead-end-pore exchange and backbone migration could shift. The literature in recent years has built the tools needed to characterize this chaotic stretching \citep{Borgne13,Borgne15,Villermaux19,Dentz23}; bringing them together with diffusiophoretic transport is, to our knowledge, an unexplored problem. 

\paragraph{Complex electrolytes and non-electrolytes.} The scalings of Sections~\ref{sec:deadend} and~\ref{sec:flow} were obtained for simple symmetric electrolytes. Multicomponent and multivalent electrolytes, and non-electrolyte solutes such as sugars, polymers, and surfactants, introduce additional gradients, competing diffusion potentials, and concentration-dependent mobilities \citep{Anderson82,Chiang14,Gupta19,Wilson20,Williams20,Yang24,Shah25,Sambamoorthy25}. How robust the macroscopic picture remains when several solutes act at once, or when $\Gamma_p$ varies strongly with composition and can reverse sign, is an important question for any real subsurface or biological setting.

\paragraph{Reactive transport.} Reactions generate gradients, and gradients drive phoresis, so reactive transport in porous media closes a loop that this chapter has largely treated as externally imposed. Sharp gradients created by fast reactions, and the pH changes that accompany them, can both drive colloid transport and alter the mobility through their effect on surface charge \citep{Shi16,Shim22b,Coleman25}. Understanding how the scaling laws survive when the gradient is reaction-generated, transient, and coupled to the evolving surface chemistry is a natural next step.

\paragraph{Confinement and tight pores.} Confinement introduces several distinct limits. If a pore throat is comparable to or smaller than the particle diameter, size exclusion, straining or steric hindrance dominate and the particle may not enter at all. If the pore remains larger than the particle but the particle is close to a wall, hydrodynamic interactions, wall diffusioosmosis and modified solute fields can change the apparent mobility; related confinement effects have been computed and measured for particles in channels and gels \citep{Doan21,Babayekhorasani16b,Phillips90,Marbach19,Ault19,Shim22,Liu25,Sambamoorthy23,Shi25,Sambamoorthy25}. 

\paragraph{Collective effects, clogging and permeability evolution.} The transport picture developed here treats the pore geometry as fixed and assumes the dilute limit, where particle--particle and particle--wall interactions are ignored. Yet diffusiophoretic drift can focus particles, raise local volume fraction, and trigger interactions that are absent for isolated tracers. At that point particles can perturb the solute and flow fields, interact through long-ranged chemical gradients, aggregate, demix, or behave more like an active suspension than a passive tracer cloud \citep{brady_2021,Wang20,Bishop23,Nasouri20,Saha20,Canalejo19}. In porous media these collective effects couple directly to clogging: deposition and salt-driven aggregation reduce permeability, redistribute flow among preferential pathways, and in extreme cases clog throats, while gradient-driven release can remobilize previously retained particles \citep{Bizmark20,Gerber2019,Wu24,Wu25}. Solute concentration also sets attachment probability through local ionic strength while its gradient sets drift (Section~\ref{sec:classical_colloid}); therefore, diffusiophoresis offers a chemical handle on where material is deposited and how the pore space evolves. This opens applications such as directed deposition for functional coatings, gradient-assisted filtration and salt-front mobilization in enhanced oil recovery \citep{Shin18,Park21,Williams20}, and may also explain protective passivation layers formed by aggregating colloids around dissolving minerals \citep{Roman25}. A predictive two-way coupling between solute gradients, evolving retention and permeability change remains to be worked out. Concretely, this means allowing the deposition rate, straining efficiency and detachment kinetics of classical retention closures (Section~\ref{sec:classical_colloid}) to depend on the local solute gradient, not only on ionic strength through the local concentration.

\paragraph{Multiphase and partially saturated porous media.} Multiphase flow is related to, but distinct from, the emulsion problem of Section~\ref{sec:emulsions}. In a partially saturated medium, the pore space available to the wetting phase is set by air--water or oil--water interfaces that may be pinned, mobile or intermittently rearranged. These interfaces create retention sites for particles, support thin films, alter hydraulic connectivity and change the velocity distribution. The broader particle-transport literature under multiphase flow emphasizes film straining, attachment to fluid--fluid interfaces, interface-mediated migration, aggregation, capillary trapping and feedback between particle accumulation and displacement patterns \citep{Bradford08,Yang2026Multiphase}. Recent simulations have begun to examine diffusiophoresis in partially saturated porous media by varying the saturation-dependent pore architecture and following colloid breakthrough \citep{Jotkar24b}. However, we still lack a fully coupled treatment of a dynamic partially saturated medium: colloid--interface interactions, moving menisci, transient permeability changes, capillary redistribution and surfactant-dependent interfacial stresses remain to be incorporated. The experimental opportunity is clear. Two-dimensional micromodels can be prepared with controlled trapped gas or oil phases, allowing salinity or surfactant gradients to be imposed while observing particles, droplets and menisci directly. Three-dimensional experiments, using refractive-index-matched packings or micro-CT-compatible systems, would then be needed to determine how these mechanisms survive in realistic pore connectivity.

\paragraph{Diffusiophoresis through complex and viscoelastic media.} Experiments in collagen gels and biofilms, Brinkman-medium theories, non-adsorbing polymer solutions and studies of surfactant/polymer gradients already show that macromolecular environments can hinder, redirect or even reverse diffusiophoretic transport \citep{Doan21,Somasundar23,Sambamoorthy23,Sambamoorthy25,Sear17,Yang23,Yang24}. What remains much less developed is the coupled transport problem in genuinely non-Newtonian flows, where viscoelastic stresses, shear-thinning viscosity or elastic instabilities reshape the solute gradients that drive mobility \citep{Aramideh19,Haward21,Kumar22,Kumar23,Browne24}. A useful next step is therefore to separate, and then recombine, two questions: how polymers and gels change the interfacial mobility, and how non-Newtonian pore-scale flows change the gradient statistics that drive that mobility.

\paragraph{Transport in biological and crowded media.} Finally, in cells, tissues, and biofilms, crowding and steric interactions with the surrounding matrix modulate the transport of colloids, cargo, and biomacromolecules \citep{Sear19,Ramm21,Doan21,Somasundar23,Sambamoorthy23,Shim24,Jambon24,Shandilya24}. A particularly intriguing implication is for intracellular organization: if gradients can systematically move macromolecules and condensates, diffusiophoresis may contribute to the spatial control of phase separation and compartmentalization within cells \citep{Alessio23,Hafner24,Doan24}. The open problem is to identify when the porous-media concepts of residence times, confinement and gradient statistics remain useful in living, crowded and self-organizing environments, and when biochemical specificity or active remodeling makes a different coarse-grained description necessary.

\paragraph{Soft-earth geophysics and soil microstructure.} A broader, largely unexplored direction is to connect gradient-driven colloid rearrangements to the soft-matter physics of soils, sediments and near-surface geophysical flows \citep{Jerolmack19,Voigtlnder24}. Soils are not passive pore networks: clay particles, organic colloids, microbes, biofilms and microplastics can aggregate, bridge grains, alter contacts and change pore-throat statistics. If chemical gradients focus or disperse these constituents, they may influence hydraulic conductivity, cohesion, yield stress, erosion thresholds, shrink--swell behavior and the transition between jammed and flowing states. This would extend diffusiophoretic porous-media transport from a problem of where particles go to a problem of how chemically driven rearrangement modifies the material properties of the ground itself, with possible implications for soil rheology, contaminant remobilization, mudflows, landslides and biologically mediated stabilization \citep{erktan2020,Pahlavan24,Rillig24,Pradeep24,Pradeep26}.

\section*{Acknowledgements}
I would like to thank members of SoFLivMat group at Yale for their contributions to the works reviewed here. I also thank participants in the 2026 KITP program on Soft Earth Geophysics for insightful discussions. I acknowledge American Chemical Society Petroleum Research Fund (65684-DNI9), Office of the Under Secretary of Defense for Research and Engineering (FA9550-22-1-0320), and National Science Foundation CAREER award (2443484) for partial support of this work.

\bibliography{references}

\end{document}